\documentclass[12pt]{article}

\usepackage{latexsym}
\usepackage{bbm}
\usepackage{amsopn,amscd}
\usepackage{amsfonts}
\usepackage{amssymb}
\usepackage{amsthm}
\usepackage{amsmath}
\usepackage{epsfig,graphics}
\usepackage{t1enc}
\usepackage[cp1250]{inputenc}
\usepackage{a4wide}

\newtheorem{Theorem}{Theorem}[section]

\newtheorem{Lemma}[Theorem]{Lemma}

\theoremstyle{proof}
\newtheorem{Remark}[Theorem]{Remark}

\newcommand{\comment}[1]{}
\newcommand{\pha}{{\mc P}}
\newcommand{\cfg}{{\mc X}}
\newcommand{\rad}{{\mc Y}}
\newcommand{\tpha}{\tilde{\mc P}}
\newcommand{\tcfg}{\tilde{\mc X}}
\newcommand{\trad}{\tilde{\mc Y}}
\newcommand{\fdom}{{\mc A}}  
\newcommand{\mc}[1]{\mathcal{#1}}
\newcommand{\mr}[1]{\mathrm{#1}}
\newcommand{\SU}{\mr{SU}}
\newcommand{\su}{\mr{su}}

\newcommand{\pr}{{\mr{pr}}}
\renewcommand{\Re}{{\mr{Re}}}
\renewcommand{\Im}{{\mr{Im}}}
\newcommand{\ol}[1]{\overline{#1}}
\newcommand{\tr}{\mathrm{tr}}

\newcommand{\RR}{\mathbb{R}}
\newcommand{\CC}{\mathbb{C}}
\newcommand{\ZZ}{\mathbb{Z}}
\newcommand{\II}{\mathbbm{1}}
\newcommand{\ctg}{\mr T^\ast}
\newcommand{\tg}{\mr T}
\newcommand{\diag}{{\rm diag}}

\newcommand{\mf}{\mathfrak}

\newcommand{\Ad}{{\mr{Ad}}}

\newcommand{\rref}[1]{{\rm \ref{#1}}}
\newcommand{\vp}{\varphi}

\newcommand{\ble}{\begin{Lemma}}
\newcommand{\ele}{\end{Lemma}}
\newcommand{\btm}{\begin{Theorem}}
\newcommand{\etm}{\end{Theorem}}
\newcommand{\bre}{\begin{Remark}\rm}
\newcommand{\ere}{\end{Remark}}
\newcommand{\beq}{\begin{equation}}
\newcommand{\eeq}{\end{equation}}
\newcommand{\beqa}{\begin{eqnarray}}
\newcommand{\eeqa}{\end{eqnarray}}
\newcommand{\linie}[3]{\put(#1){\line(#2){#3}}}
\newcommand{\marke}[3]{\put(#1){\put(0.05,0.1){\makebox(-0.1,-0.2)[#2]{$#3$}}}}

\begin{document}

\title{\bf A Lattice Gauge Model of Singular Marsden-Weinstein Reduction\\ Part I.
Kinematics}

\author{
E.~Fischer$^1$, G.~Rudolph$^2$, M.~Schmidt$^{2,3}$,
\\[0.3cm]
$^1$ Max-Planck Institute for the Physics of Complex Systems,
\\
N\"othnitzer Stra{\ss}e 38, 01187 Dresden, Germany
\\[0.3cm]
$^2$ Institute for Theoretical Physics, University of Leipzig\\
Augustusplatz 10/11, 04109 Leipzig, Germany
\\[0.3cm]
$^3$ Corresponding author
\\
e-mail:~ matthias.schmidt@itp.uni-leipzig.de
\\
phone:~ +49-341-9732431,~
fax:~ +49-341-9732548
 }

\maketitle

\begin{abstract}

The simplest nontrivial toy model of a classical $\SU(3)$ lattice gauge theory
is studied in the Hamiltonian approach. \comment{Here, the lattice consists of a
single plaquette. For this model, t}%
By means of singular symplectic reduction, the reduced phase space is
constructed. Two equivalent descriptions of this space in terms of a symplectic
covering as well as in terms of invariants are derived.

\end{abstract}

{\bf MSI:}~ 37J15, 70S15
\\[0.3cm]
{\bf Subject classification:}~ symplectic geometry, classical mechanics
\\[0.3cm]


\section{Introduction}


In the study of quantum gauge theory by nonperturbative methods there exist, in
effect, two approaches: one is to quantize the unreduced system and then reduce
the symmetries on the quantum level, the other one is to first reduce the
symmetries on the classical level and then quantize the reduced system.
For a discussion of the first strategy within the framework of lattice gauge
theory, see \cite{KiRu:JMP} and the references therein. The aim of the
present paper is to contribute to the second approach. The
motivation behind stems from the well-known fact that nonabelian gauge fields
can have several symmetry types, which give rise to singularities in the
'true' (i.e., reduced) classical configuration space. Mathematically spoken, the
latter is a stratified space rather than a smooth manifold. It is natural to ask
whether the singularities produce physical effects. For a systematic study of
this open problem one needs a concept of how to implement the singularity
structure in quantum theory. Such concepts have been developed
in recent years, see, e.g., \cite{Hue1,Hue2,SGQ}.
To separate the problem of symmetry reduction from
the usual problems of a field theory related to the infinite number of
degrees of freedom, it is reasonable to first study lattice gauge theory. This way, one obtains a variety of toy models to form and test concepts
and to develop quantum theory on a space with singularities.
It is important for quantum theory, as well as interesting in its own right,
to understand the classical dynamics of these models. Thus, in the present
paper we consider the simplest nontrivial model of an $\SU(3)$ lattice gauge
theory, where the lattice consists of a single plaquette. We study the
kinematics of this model, i.e., the structure of the reduced phase space.
The classical dynamics will then be studied in a subsequent paper.

We proceed as follows. In Section \rref{Smodel} we introduce the model. In
Section \rref{Ssymred} we carry out symmetry reduction. This
will lead  us to the so-called reduced cotangent bundle
\cite{LermanMontgomerySjamaar}. Then we give two
equivalent descriptions of this bundle. One is in terms of a symplectic covering
(Section \rref{Ssplcov}), the other one is in terms of invariants (Section
\rref{Sivr}). We conclude with some general remarks on the dynamics in Section
\rref{Sdynamics}.


\section{The model}
\label{Smodel}


Let us consider chromodynamics on a finite regular
cubic lattice $\Lambda $. Denote the set of the oriented, $i$-dimensional
elements of $\Lambda$ by ${\Lambda}^i$ (sites, links, plaquettes and cubes
in increasing order of $i$).
The gauge group is $G=\SU(3)$, its Lie algebra is ${\mathfrak g}=\su(3)$.
The classical gluonic potential is approximated by its parallel transporter:
$$
{\Lambda}^1 \ni (x,y) \mapsto a_{(x,y)} \in G \, .
$$
Thus, the unreduced classical configuration space is the direct product
$G^{\Lambda^1}$ and the unreduced phase space is $\ctg
G^{\Lambda^1}$. By means of the natural isomorphisms $\ctg G^{\Lambda^1} \cong (\ctg
G)^{\Lambda^1}$, $\ctg G\cong G\times\mf g^\ast$ and $\mf g^\ast\cong\mf g$, see
below for details, the canonically conjugate momenta (colour electric fields)
are given by maps
$$
{\Lambda}^1 \ni (x,y) \mapsto A_{(x,y)} \in {\mathfrak g} \, .
$$
Local gauge transformations are approximated by maps
$$
\Lambda^0 \ni x \mapsto g_x \in G\,,
$$
hence the group of local gauge transformations is the direct product
$G^{\Lambda^0}$. It acts on the phase space as follows:
$$
a_{(x,y)}^{\prime}  =  g_x \cdot a_{(x,y)} \cdot g_y^{-1}
 \,,~~~~~~
A_{(x,y)}^{\prime}  =  g_x \cdot A_{(x,y)} \cdot g_x^{-1} \,.
$$
The (gauge invariant) Hamiltonian is given by
 \beq
 \label{GHgeneral}
 \begin{array}{c}
H
 =
 -
\frac{\delta^3}{2}  \sum_{(x,y) \in \Lambda^1} \tr\big(A_{(x,y)}^2\big)
 +
\frac{1}{2 g^2 \delta} \sum_{p \in \Lambda^2}
 \left(
6 - \tr(a_p + a_p^{\dagger})
 \right)\,.
 \end{array}
 \eeq
Here, $\delta$ and $g$ denote the lattice spacing and the coupling constant,
respectively, and $a_p$ is the parallel transporter around the plaquette
$p$. For a plaquette $p$ with vertices $x,y,z,u$ we choose
$$
a_p = a_{xy}\cdot a_{yz}\cdot a_{zu}\cdot a_{ux}\,.
$$
While $a_p$ depends on the choice of a base point $x$, $\tr(a_p)$ does not.

In the present paper we consider the case where $\Lambda$ consists of a single
plaquette. This is the simplest nontrivial model for a Hamiltonian lattice
gauge theory. On three of the links of the plaquette, $a$ and $A$ can be gauged
to $\II$ and $0$, respectively. Such a gauge is called a tree gauge. Then the
residual gauge freedom consists of
constant gauge transformations. Thus, the unreduced configuration space is the
group manifold $G$ and the unreduced phase space is $\ctg G\cong G\times\mf g$.
Its elements will be denoted by $(a,A)$. The gauge group is $G$, its action on
the phase space is given by diagonal conjugation
$$
a' = g a g^{-1}
 \,,~~~~~~
A' = g A g^{-1}\,.
$$
The Hamiltonian becomes
 \beq
 \label{GH}
 \begin{array}{c}
H
 =
 -
\frac{\delta^3}{2}  \tr(A^2)
 +
\frac{1}{2 g^2 \delta} \left(6 - \tr(a + a^{\dagger}) \right)\,.
 \end{array}
 \eeq

Next, we will carry out symmetry reduction. The basic object for
this is the $G$-manifold of the unreduced configuration space,
because it determines the kinematical structure of the model
completely.


\section{Symmetry reduction}
\label{Ssymred}


First, let us recall the general procedure. It is known as
cotangent bundle reduction and is a special case of (singular)
Marsden-Weinstein reduction.


\subsection{Cotangent bundle reduction}
\label{SScotabunred}


Let $Q$ be a manifold acted upon properly by a Lie group $K$ (we
may even assume that $K$ is compact). Associated with $(Q,K)$
there is the surjection
 \beq\label{Gredcotabungeneral}
\pi : \ctg (Q/K)\to Q/K\,.
 \eeq
The base space $Q/K$ consists of the $K$-orbits in $Q$, equipped with the
quotient topology, the stratification by the orbit types of $K$-action and the
smooth structure
$$
C^\infty(Q/K) := C^\infty(Q)^K
$$
(invariant smooth functions on $Q$). Thus, $Q/K$ is a stratified
topological space with smooth structure, see \cite{Pflaum:LNM} for
this notion.

The total space $\ctg(Q/K)$ is obtained as follows. The action of $K$ on $Q$
lifts to a proper symplectic action of $K$ on the cotangent bundle $\ctg Q$ by
the corresponding point transformations. The map $J:\ctg Q \to \mf k^\ast$
defined by
 \beq\label{Gdefmomapgeneral}
\langle J(\eta),X \rangle := \eta(X^Q)
 \,,~~~~~~
\eta\in\ctg Q\,,~~X\in\mf k\,,
 \eeq
where $X^Q$ denotes the Killing vector field associated with $X$,
is an equivariant momentum mapping for this action
\cite[\S 4.2]{AbraMar}. (Thus, these data define a Hamiltonian
$G$-manifold naturally associated with $(Q,K)$.) Since $J$ is
equivariant, the level set $J^{-1}(0)$ is invariant under $K$. The
bundle space $\ctg(Q/K)$ is given by the topological quotient
$J^{-1}(0)/K$. It is equipped with the following structure, see
\cite{ArmsCushmanGotay,OrtegaRatiu,SjamaarLerman} or \cite[App.\ B.5]{CuBa}:
\medskip

-- A smooth Poisson structure. The natural smooth structure of
$\ctg(Q/K)$ is given by
$$
C^\infty\big(\ctg(Q/K)\big) := C^\infty(\ctg Q)^K/V^K\,,
$$
where $V$ denotes the vanishing ideal of the level set
$J^{-1}(0)$. Since $K$ acts symplectically on $\ctg Q$,
$C^\infty(\ctg Q)^K$ is a Poisson subalgebra of $C^\infty(\ctg Q)$. In view of
Noether's theorem, $J^{-1}(0)$ is invariant under the Hamiltonian flow of
invariant functions. Hence, $V^K$ is a Poisson ideal in $C^\infty(\ctg Q)^K$.
Therefore, $C^\infty\big(\ctg(Q/K)\big)$ inherits a Poisson bracket through
$$
\{f+V^K,g+V^K\}_{\ctg(Q/K)} = \{f,g\}_{\ctg Q}
 \,,~~~~~~
f,g\in C^\infty\big(\ctg(Q/K)\big)\,.
$$
\medskip

-- A stratification by orbit types. Using the slice theorem it can
be shown that for given orbit type $\tau$ the subset
$J^{-1}(0)_\tau$ of $J^{-1}(0)$ consisting of the elements of type
$\tau$ is an embedded submanifold of $\ctg Q$. Local charts on the
$\tau$-stratum $\ctg(Q/K)_\tau$ of $\ctg(Q/K)$ are then defined in
the usual way: for a given point in $\ctg(Q/K)_\tau$ one chooses a
representative in $J^{-1}(0)_\tau$ and a slice about the
representative for the action of $K$ on $J^{-1}(0)_\tau$. By
restriction, the natural projection $\pi_\tau:J^{-1}(0)_\tau \to
\ctg(Q/K)_\tau$ induces a homeomorphism of the slice onto its image. Thus,
charts on the slice induce charts on $\ctg(Q/K)_\tau$.
\medskip

-- Symplectic structures on the strata $\ctg(Q/K)_\tau$. One can
prove that the annihilator of the pull-back of the symplectic form
$\omega$ of $\ctg Q$ to the submanifold $J^{-1}(0)_\tau$ coincides
with the distribution defined by the tangent spaces of the orbits.
Therefore, the pull-back of $\omega$ to a slice for the action of
$K$ on $J^{-1}(0)_\tau$ is a symplectic form on that slice.
Through the homeomorphism of the slice onto its image in $\ctg(Q/K)_\tau$,
induced by the natural projection $\pi_\tau$ it defines a local symplectic
form on $\ctg(Q/K)_\tau$. Due to the
fact that $\omega$ is $K$-invariant, all the local forms merge to
a symplectic form $\omega^\tau$ on $\ctg(Q/K)_\tau$. Then
$$
\pi_\tau^\ast\,\omega^\tau = j_\tau^\ast\,\omega\,,
$$
where $j_\tau : J^{-1}(0)_\tau\to\ctg Q$ denotes the natural injection.
\medskip

By construction, the injections $(\ctg Q)_\tau\to\ctg(Q/K)$ are
Poisson maps. Therefore, the above data turn $\ctg(Q/K)$ into a
stratified symplectic space.
\medskip

Finally, the projection $\pi$ of \eqref{Gredcotabungeneral} is
induced by the restriction of the natural (equivariant) projection
$\ctg Q\to Q$ to the level set $J^{-1}(0)$. Since $J^{-1}(0)$
contains the zero section of $\ctg Q$, $\pi$ is surjective.

\bre\label{Rprojstrat}

The fibres of \eqref{Gredcotabungeneral} may intersect several
distinct strata of $\ctg(Q/K)$. In particular, $\pi$ does not
preserve the orbit types. However, as the stabilizer of a covector
in $\ctg Q$ cannot be larger than that of its base point, $\pi$
does not decrease orbit types. For a detailed study of the
stratifications of the fibres of $\ctg(Q/K)$, see
\cite{Perlmutter}.

\ere

\bre\label{Rterminology}

Since \eqref{Gredcotabungeneral} is a bundle in the topological
category in the sense of \cite{Husemoeller} and since it plays the
same role for $Q/K$ as the cotangent bundle $\ctg Q$ plays for
$Q$, \eqref{Gredcotabungeneral} is called the reduced cotangent
bundle in \cite{LermanMontgomerySjamaar}, although in general its
elements are not covectors. When $K$ acts freely then $Q/K$ is a
manifold and \eqref{Gredcotabungeneral} is isomorphic to the
cotangent bundle of this manifold \cite{AbraMar}. In general, the cotangent
bundles of the strata of $Q/K$ are dense subsets of the corresponding strata
of $\ctg(Q/K)$ \cite{Perlmutter}.

\ere

If, like in our case, $(Q,K)$ is the configuration space of a Hamiltonian
system with symmetries, $Q/K$ and $\ctg(Q/K)$ are referred to as the
reduced configuration space and the reduced phase space, respectively.
It can be shown in general \cite{OrtegaRatiu} that if an evolution curve
in $\ctg Q$ w.r.t.\ a $K$-invariant Hamiltonian meets a submanifold
$J^{-1}(0)_\tau$
then it is contained completely in this submanifold. Therefore, dynamics in
$\ctg(Q/K)$ takes place inside the strata. Due to Remark \rref{Rprojstrat}, an
analogous statement for $Q/K$ is in general not true, though.

We will now discuss the reduced data of our model in detail. The reduced
configuration space $Q/K$ and the reduced phase space $\ctg(Q/K)$ will be
denoted by $\cfg$ and $\pha$, respectively.


\subsection{The reduced configuration space $\cfg$}


In what follows we will write $G$ for $\SU(3)$ and $\mf g$ for
$\su(3)$.

By construction, $\cfg$ is the adjoint quotient $G/\Ad$. As $G$ is
semisimple, this space has the following two standard
realizations. Let $T$ denote the subgroup of diagonal matrices of
$G$. One has $T\cong \mr U(1)\times\mr U(1)$, a $2$-torus. For
$j=1,2,3$, let $T_{(j)}$ denote the subsets of $T$ consisting of
the elements whose entries coincide, possibly except for the $j$th
one. Let $\fdom$ denote one of the triangular subsets of $T$ which
are cut out by the $T_{(j)}$, $j=1,2,3$, see Figure
\rref{figfdom}. From the embedding $\fdom \to T$, $\fdom$ acquires
a Whitney smooth structure. It is a standard fact that the
embeddings $\fdom\to T\to G$ induce, by passing to quotients,
isomorphisms
 \beq\label{Gisocfg}
\cfg \cong T/S_3 \cong\fdom
 \eeq
of topological spaces with smooth structure. Here the symmetric
group $S_3$ acts by permutation of entries and the smooth
structure of $T/S_3$ is defined by the invariant smooth functions
on $T$.

 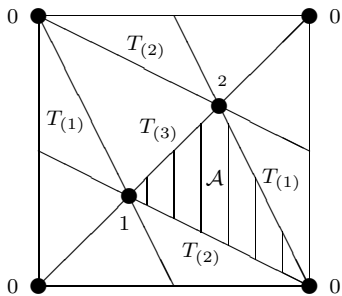
\begin{figure}

 \center

\unitlength1.8cm

 \begin{picture}(2,2.1)
\put(0,0.1){
 \linie{0,1}{2,-1}{2}
 \linie{0,2}{2,-1}{2}
 \linie{0,2}{1,-2}{1}
 \linie{1,2}{1,-2}{1}
 \linie{0,0}{1,1}{2}
 \linethickness{0.15pt}
 \linie{0,0}{0,1}{2}
 \linie{2,0}{0,1}{2}
 \linie{0,0}{1,0}{2}
 \linie{0,2}{1,0}{2}
 \put(0,0){\circle*{0.12}}
 \marke{-0.1,0}{cr}{\scriptstyle 0}
 \put(2,0){\circle*{0.12}}
 \marke{2.1,0}{cl}{\scriptstyle 0}
 \put(0.6666,0.6666){\circle*{0.12}}
 \marke{0.6333,0.6166}{tc}{\scriptstyle 1}
 \put(1.3333,1.3333){\circle*{0.12}}
 \marke{1.3566,1.3833}{bc}{\scriptscriptstyle 2 }
 \put(0,2){\circle*{0.12}}
 \marke{-0.1,2}{cr}{\scriptstyle 0}
 \put(2,2){\circle*{0.12}}
 \marke{2.1,2}{cl}{\scriptstyle 0}
 \marke{1.0866,0.9666}{br}{\scriptstyle T_{(3)}}
 \marke{0.4,1.4}{tr}{\scriptstyle T_{(1)}}
 \marke{1.6,0.6}{bl}{\scriptstyle T_{(1)}}
 \marke{0.6,1.58}{bl}{\scriptstyle T_{(2)}}
 \marke{1.4,0.42}{tr}{\scriptstyle T_{(2)}}
 \linie{0.8,0.8}{0,-1}{0.2}
 \linie{1,1}{0,-1}{0.5}
 \linie{1.2,1.2}{0,-1}{0.8}
 \linie{1.4,1.2}{0,-1}{0.9}
 \linie{1.6,0.8}{0,-1}{0.6}
 \linie{1.8,0.4}{0,-1}{0.3}
 \marke{1.3,0.8}{cc}{\scriptstyle\fdom}
 }
 \end{picture}

\caption{\label{figfdom} A possible choice for the subset $\fdom$ of $T$.
The numbers $0,1,2$ stand for the central elements $\II$,
$\mr e^{\mr i\frac{2}{3}\pi}\II$ and $\mr e^{\mr i\frac{4}{3}\pi} \II$,
respectively.}

 \end{figure}

 \comment{

 \begin{figure}
\center
\footnotesize
\unitlength1.2cm
 \begin{picture}(2,2.7)

 \put(0,0.5){
 \thicklines
 \put(0,0){\circle*{0.17}}
 \marke{0,0}{tr}{\II}
 \linie{0,0}{1,0}{2}
 \marke{1,-0.1}{tc}{T_{(3)}}
 \put(2,0){\circle*{0.17}}
 \marke{2,0}{tl}{~e^{i2\pi/3}\II}
 \linie{0,0}{3,5}{1}
 \marke{0.3,0.8333}{cr}{T_{(1)}}
 \put(1,1.6666){\circle*{0.17}}
 \marke{1,1.7}{bc}{e^{i4\pi/3}\II}
 \linie{2,0}{-3,5}{1}
 \marke{1.7,0.8333}{cl}{T_{(2)}}
 \linethickness{0.1pt}
 \linie{0.1,0}{0,1}{0.1666}
 \linie{0.2,0}{0,1}{0.3333}
 \linie{0.3,0}{0,1}{0.5}
 \linie{0.4,0}{0,1}{0.6666}
 \linie{0.5,0}{0,1}{0.8333}
 \linie{0.6,0}{0,1}{1}
 \linie{0.7,0}{0,1}{1.1666}
 \linie{0.8,0}{0,1}{1.3333}
 \linie{0.9,0}{0,1}{1.5}
 \linie{1,0}{0,1}{1.6666}
 \linie{1.9,0}{0,1}{0.1666}
 \linie{1.8,0}{0,1}{0.3333}
 \linie{1.7,0}{0,1}{0.5}
 \linie{1.6,0}{0,1}{0.6666}
 \linie{1.5,0}{0,1}{0.8333}
 \linie{1.4,0}{0,1}{1}
 \linie{1.3,0}{0,1}{1.1666}
 \linie{1.2,0}{0,1}{1.3333}
 \linie{1.1,0}{0,1}{1.5}
 }

\end{picture}

\caption{\label{figfdom} A possible choice for the subset $\fdom$ of $T$. The
other cases are obtained by permutation of either the vertices or the edges.}

 \end{figure}
 }

Let us describe the stratification. The number of distinct entries
of $a\in \fdom$ can be $3$, $2$ or $1$. Denote the corresponding
subsets of $\fdom$ by $\fdom_k$ with  $k=2,1,0$. One has $\fdom_1
= \bigcup_{j=1}^3 \fdom\cap T_{(j)}$. Topologically, $\fdom$ is a
$2$-simplex, $\fdom_2$ is its interior, $\fdom_1$ consists of the
edges without the vertices and $\fdom_0$ consists of the vertices.
Taking into account that the stabilizer of $a$ under the action of
$\SU(3)$ is given by the centralizer of $a$ in $\SU(3)$, the
stabilizer of $a\in\fdom_k$ is
 \beq\label{Gstabcfg}
\begin{array}{c|c|c}
k
 &
\SU(3)\text{-stabilizer}
 &
S_3\text{-stabilizer}
\\ \hline
2 & T & \{\II\}
\\
1 & \mr U(2) & S_2
\\
0 & \SU(3) & S_3
\end{array}
 \eeq
In particular, $\fdom_0 = \ZZ_3$, the centre of $\SU(3)$. Denote
the orbit types in the respective order by $\tau_2$, $\tau_1$ and
$\tau_0$, irrespective of the action they belong to, and the
corresponding strata of $\cfg$ by $\cfg_2$, $\cfg_1$ and $\cfg_0$.
(The numbering refers to the dimensions of the strata.) Type
$\tau_2$ is the principal orbit type and $\cfg_2$ is the principal
stratum.

It is easy to see that the isomorphism $\cfg\cong\fdom$ holds on
the level of stratified smooth topological spaces.

\bre

The identification of  $\cfg$ with $\fdom$ endows $\cfg$ with a
CW-complex structure in an obvious fashion. Already for the
quotient $\big(\SU(3)\times\SU(3)\big)/\SU(3)$ with $\SU(3)$
acting by diagonal conjugation, which is the reduced configuration
space of lattice $\SU(3)$-gauge theory on a lattice with $2$
plaquettes, the construction of a CW-complex structure is much
more complicated, see \cite{CRS}.

\ere


\subsection{The reduced phase space $\pha$}
\label{SSredphaspa}


As anticipated in Section \rref{Smodel}, we identify $\ctg G$ with
the direct product
$G\times\mf g$ by virtue of the natural diffeomorphism
 \beq\label{Gdefisocotabun}
G\times\mf g \to \ctg G
 \,,~~~~~~
(a,A)\mapsto \langle A,\mr R_{a^{-1}}'\,\cdot\,\rangle\,.
 \comment{
\\ \label{Gdefisotabun}
G\times\mf g & \to & \tg G
 \,,~~~~~~
(a,A)\mapsto \mr R_a' A\,.
 }
 \eeq
Here, $\mr R_a : G\to G$ denotes right multiplication by $a\in G$
and $\langle\,\cdot\,,\,\cdot\,\rangle$ is the ordinary scalar product of
complex matrices,
$$
\langle A,B \rangle = \tr(A^\dagger B)
 \,,~~~~~~
A,B\in\mr M_3(\CC)\,.
$$
When restricted to $\mf g$ this form yields a real scalar product
which, up to a constant factor, coincides with the negative of the
Killing form of $\mf g$:
$$
\langle A,B \rangle = - \tr(AB)
 \,,~~~~~~
A,B\in\mf g\,.
$$
Since $\tg(G\times\mf g) \cong \tg G \times \tg\mf g$, vectors
tangent to $G\times\mf g$ at $(a,A)$ can be written as $(\mr
R_a'B,(A,C))$ with $B,C\in\mf g$. Under the identification
\eqref{Gdefisocotabun} the symplectic potential of $\ctg G$ takes
the standard form
 \beq
 \label{Gtheta}
\theta_{(a,A)}\big(R_a'B,(A,C)\big) = \langle A , B \rangle\,,
 \eeq
hence the symplectic form $\omega = d \theta$ is
 \beq
 \label{Gomega}
\omega_{(a,A)}\Big(\,(R_a'B_1,(A,C_1))\,,\,(R_a'B_2,(A,C_2))\,\Big)
 =
\langle B_1,C_2\rangle - \langle C_1,B_2\rangle
 -
\langle A,[B_1,B_2]\rangle\,.
 \eeq
The action of $G$ on $\ctg G$ by the induced point transformations
is given by conjugation, i.e.,
 \beq
 \label{Gactioncotabun}
b\cdot(a,A) = \left(b\,a\,b^{-1}\,,\,b\,A\,b^{-1}\right)\,.
 \eeq
If we furthermore identify $\mf g^\ast$ with $\mf g$ by virtue of the
scalar product $\langle\,\cdot\,,\,\cdot\,\rangle$, the natural
momentum mapping for this action is given by the map
 \beq
 \label{Gmomap}
J:G\times\mf g\to \mf g
 \,,~~~~~~
J(a,A) = A - a^{-1} A a\,.
 \eeq
The level set $J^{-1}(0)$ is therefore given by all pairs
$(a,A)\in G\times \mf g$ where $a$ and $A$ commute. In particular,
it contains the subset $T\times\mf t$. By restriction of the
natural projection to orbits we obtain a map
 \beq\label{GdefcovT}
\lambda : T\times\mf t \to \pha\,.
 \eeq
Let $(a,A)\in J^{-1}(0)$. Since $a$ and $A$ commute, they possess
a common eigenbasis. Since $a$ is unitary and $A$ is
anti-Hermitian, the eigenbasis can be chosen to be orthonormal.
Hence, by $G$-action, $(a,A)$ can be transported to $T\times\mf
t$. In other words, every $G$-orbit in $J^{-1}(0)$ intersects the
subset $T\times\mf t$. Hence, $\lambda$ is surjective. Since two
elements of $T\times\mf t$ are conjugate under $G$ iff they differ
by a simultaneous permutation of their entries, then $\lambda$
descends to a bijection
$$
(T\times\mf t)/S_3 \to \pha\,.
$$
Standard arguments ensure that this is in fact a homeomorphism.
Thus, we can use $\lambda$ to describe $\pha$. In particular,
$\pha$ is an orbifold.

We start with the stratification. The number of entries which
simultaneously coincide for both $a$ and $A$ can be $0$, $2$ or
$3$. Denote the corresponding subsets of $T\times\mf t$ by
$(T\times\mf t)_k$ with $k=2,1,0$, respectively. The stabilizers
and orbit types of $(a,A)\in(T\times\mf t)_k$ under
$\SU(3)$-action and $S_3$-action are
 \beq\label{Gstabpha}
\begin{array}{c|c|c|c}
k
 &
\SU(3)\text{-stabilizer}
 &
S_3\text{-stabilizer}
 &
\text{orbit type}
\\ \hline
2 & T & \{\II\} & \tau_2
\\
1 & \mr U(2) & S_2 & \tau_1
\\
0 & \SU(3) & S_3 & \tau_0
\end{array}
 \eeq
Since the orbit types are the same as for $\cfg$ we use the same
notation. Let $\pha_k\subseteq\pha$ denote the stratum of type
$\tau_k$, $k=0,1,2$. $\pha_2$ is the principal stratum. Since the subsets
$(T\times\mf t)_k$ are the pre-images of the strata $\pha_k$ under $\lambda$,
they will be referred to as strata of $T\times\mf t$. By
restriction, $\lambda$ induces maps
 \beq\label{GdefcovTk}
\lambda_k : (T\times\mf t)_k \to \pha_k
 \,,~~~~~~
k=2,1,0\,,
 \eeq
which descend to homeomorphisms of $(T\times\mf t)_k/S_3$ onto
$\pha_k$, $k=2,1,0$.

We determine $(T\times\mf t)_k$ explicitly. Recall that $\ZZ_3$
denotes the center of $G=\SU(3)$. As for $T_{(j)}$, let $\mf
t_{(j)}$, $j=1,2,3$, denote the subset of $\mf t$ consisting of
the elements whose entries coincide, possibly except for the $j$th
one. We find
$$
 \begin{array}{rcl}
(T\times\mf t)_0 & = & \ZZ_3\times\{0\}\,,
\\
(T\times\mf t)_1
 & = &
 \left(
\bigcup\nolimits_{j=1}^3 T_{(j)} \times \mf t_{(j)}
 \right)
 -
(T\times\mf t)_0\,,
\\
(T\times\mf t)_2 & = & T\times\mf t - (T\times\mf t)_1\,.
 \end{array}
$$
These are embedded submanifolds of $T\times\mf t$. Since $\mf t$
is the Lie subalgebra of $\mf g$ associated with the Lie subgroup
$T$ of $G$, $T\times \mf t$ is a {\em symplectic} submanifold of
$G\times\mf g$. Analogously, so are $T_{(j)}\times\mf t_{(j)}$,
$j=1,2,3$. It follows that $(T\times\mf t)_k$, $k=2,1$, are
symplectic manifolds. For convenience, in the following we will
view $(T\times\mf t)_0$ as a (trivially) symplectic manifold, too.

\btm\label{TphaTt}

The map $\lambda$ is Poisson. The maps $\lambda_k$ are local
symplectomorphisms.

\etm

{\it Proof.}~ By definition, $C^\infty(\pha)$ is a quotient of
$C^\infty(G\times\mf g)^G$. Hence, the first assertion is a direct
consequence of the fact that $T\times\mf t$ is a symplectic
submanifold of $G\times \mf g$. For the second assertion, recall
the construction of the symplectic forms on the strata $\pha_k$
from Subsection \rref{SScotabunred}. The assertion then follows by
observing that any point of $\pha_k$ has a representative in
$(T\times\mf t)_k$ and that a sufficiently small neighbourhood of
the chosen representative in $(T\times\mf t)_k$ provides a slice
for the action of $G$ on the submanifold $J^{-1}(0)_k$ of
$G\times\mf g$. Here $J^{-1}(0)_k$ denotes the subset of
$J^{-1}(0)$ consisting of the elements of the orbits of type
$\tau_k$.
 \qed

\bre

1.~ Since the submanifolds $(T\times\mf t)_k$ are symplectic and since
$S_3$ is finite, the quotient $(T\times\mf t)/S_3$ naturally
carries the structure of a stratified symplectic space. Of course,
this structure might be viewed as to be obtained by singular
Marsden-Weinstein reduction with (necessarily) trivial momentum
map. Then Theorem \rref{TphaTt} says that the map $\lambda$
induces an isomorphism of stratified symplectic spaces of
$(T\times\mf t)/S_3$ onto $\pha$.
\medskip

2.~ Dynamics on $\pha$ is thus given by the dynamics on
$T\times\mf t$ w.r.t.\ an $S_3$-invariant Hamiltonian and the
symplectic form \eqref{Gomega}. Similarly, motion on $\cfg$ is
given by $S_3$-invariant motion on the $2$-torus with metric
defined by the scalar product $\langle\,\cdot\,,\,\cdot\,\rangle$.

\ere


\subsection{The projection $\pi:\pha\to\cfg$}
\label{SSredcotabun}

\begin{figure}
 \centering
\unitlength1.6cm
 \begin{picture}(8,2)
 \put(1,1.15){
\linie{0,0}{0,1}{0.6}
\linie{0,0}{0,-1}{0.6}
\linie{0,0}{5,3}{1}
\linie{0,0}{-5,-3}{1}
\linie{0,0}{5,-3}{1}
\linie{0,0}{-5,3}{1}
\marke{0,0.6}{bc}{\scriptstyle \mf t_1}
\marke{-1,0.6}{bc}{\scriptstyle \mf t_2}
\marke{-1,-0.6}{tc}{\scriptstyle \mf t_3}
\linethickness{0.1pt}
\multiput(-0.95,0)(0,0.095){7}{\linie{0,0}{1,0}{1.9}}
\multiput(-0.95,0)(0,-0.095){7}{\linie{0,0}{1,0}{1.9}}
\marke{0,-0.75}{tc}{a\in\cfg_2}
 }
 \put(4,1.15){
 {
\thicklines
\linie{0,0}{0,1}{0.6}
\linie{0,0}{0,-1}{0.6}
 }
\linie{0,0}{5,3}{1}
\linie{0,0}{-5,-3}{1}
\linie{0,0}{5,-3}{1}
\linie{0,0}{-5,3}{1}
\marke{0,0.6}{bc}{\scriptstyle \mf t_1}
\marke{-1,0.6}{bc}{\scriptstyle \mf t_2}
\marke{-1,-0.6}{tc}{\scriptstyle \mf t_3}
\linethickness{0.1pt}
\multiput(0,0)(0,0.095){7}{\linie{0,0}{1,0}{0.95}}
\multiput(0,0)(0,-0.095){7}{\linie{0,0}{1,0}{0.95}}
\marke{0,-0.75}{tc}{a\in\cfg_1~~(a\in T_{(1)})}
 }
 \put(7,1.15){
 {
\thicklines
\linie{0,0}{0,1}{0.6}
\linie{0,0}{5,3}{1}
 }
\linie{0,0}{0,-1}{0.6}
\linie{0,0}{-5,-3}{1}
\linie{0,0}{5,-3}{1}
\linie{0,0}{-5,3}{1}
\marke{0,0.6}{bc}{\scriptstyle \mf t_1}
\marke{-1,0.6}{bc}{\scriptstyle \mf t_2}
\marke{-1,-0.6}{tc}{\scriptstyle \mf t_3}
\linethickness{0.1pt}
\linie{0,0.095}{1,0}{0.15827}
\linie{0,0.19}{1,0}{0.31654}
\linie{0,0.285}{1,0}{0.475}
\linie{0,0.38}{1,0}{0.63308}
\linie{0,0.475}{1,0}{0.79135}
\linie{0,0.57}{1,0}{0.95}
\marke{0,-0.75}{tc}{a\in\cfg_0}
 }
 \end{picture}

\caption{\label{figfibres} The fibres $\pi^{-1}(a)$.}

 \end{figure}
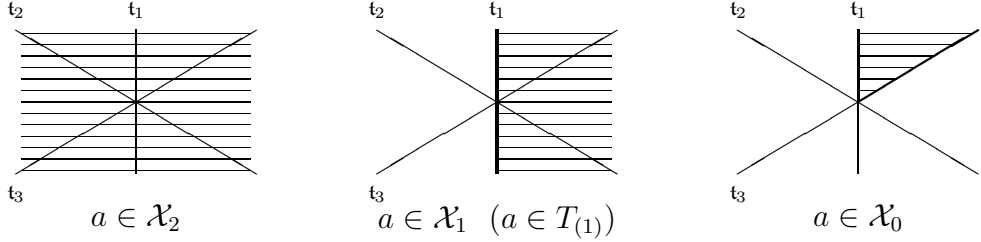

Recall from Subsection \rref{SScotabunred} that the projection 
$\pi:\pha\to\cfg$ is induced by the cotangent bundle projection $\ctg G \to G$.
By virtue of the identification \eqref{Gdefisocotabun}, the latter is identified
with the natural projection to the first factor $\pr_1:G\times\mf
g\to G$. Hence, one has the following commutative diagram
$$
\begin{CD}
T\times\mf t @>\lambda>> \pha
\\
@V \pr_1 VV @VV \pi V
\\
T @>>> \cfg
\end{CD}
$$
where the lower horizontal arrow is defined by restriction of the
natural projection $G\to\cfg$. It follows that the fibre over
$a\in\cfg$ ($\cfg$ being identified with $\fdom$ and hence with a
subset of $T$) is given by
$$
\pi^{-1}(a) = \mf t/S(a)\,,
$$
where $S(a)$ is the stabilizer of $a$ under the action of $S_3$.
According to \eqref{Gstabcfg}, there are $3$ cases, illustrated in
Figure \rref{figfibres}.
\medskip

-- If $a\in\cfg_2$, $S(a)$ is trivial, hence $\pi^{-1}(a) = \mf
t$. I.e., the fibre is a full $2$-plane and belongs to the stratum
$\pha_2$.
\medskip

-- If $a\in\cfg_1$ then $a\in T_{(j)}-\ZZ_3$ for some $j=1,2,3$.
Then $S(a) = S_2$, acting by permuting the $2$ entries besides the
$j$th one. Hence, $\pi^{-1}(a) = \mf t/S_2$, acting by reflection
about the subspace $\mf t_{(j)}$. Therefore, the fibre may be
identified with one of the two half-planes of $\mf t$ cut out by
$\mf t_{(j)}$. Its interior belongs to the stratum $\pha_2$,
whereas the boundary $\mf t_{(j)}$ belongs to the stratum
$\pha_1$.
\medskip

-- If $a\in\cfg_0$, i.e., $a\in\ZZ_3$, then $S(a) = S_3$. The
action of $S_3$ on $\mf t$ is generated by the reflections about
the $3$ subspaces $\mf t_{(j)}$, $j=1,2,3$. Hence, the fibre may
be identified with one of the $6$ (closed) Weyl chambers of $\mf
t$ cut out by $\mf t_{(j)}$, $j=1,2,3$ (the walls of the Weyl chambers).
The interior of the Weyl chamber belongs to the stratum $\pha_2$,
the walls minus the origin belong to the stratum $\pha_1$ and the
origin belongs to the stratum $\pha_0$.
\medskip

One can see explicitly that the projection $\pi : \pha\to\cfg$
does not preserve the stratification, because the fibres over points in $\cfg_1$
and $\cfg_0$ intersect more than 1 stratum of $\pha$. As stated in Remark
\rref{Rprojstrat}, this is a general phenomenon.

\bre\label{RredphaspaSU(2)}

The description of the reduced data given here generalizes to
an arbitrary compact semisimple Lie group in an obvious way: $T$ and $\mf
t$ are replaced by a maximal torus in $G$ and its Lie algebra,
which is a Cartan subalgebra of $\mf g$. $\fdom$ is replaced by a Weyl alcove
in $T$ and $S_3$ is replaced by the Weyl group of $G$. It is interesting that
for $G=\SU(2)$ one obtains the reduced phase space of the spherical pendulum
with zero angular momentum, which is the well-known canoe \cite[\S VI.2]{CuBa}.

\ere

This completes the construction of the reduced data for the model
under consideration. Next, we will derive tools to study the dynamics of this
model. I.e., first, a symplectic covering of $T\times\mf t$ and, second, a
description of $\pha$ and $\cfg$ in terms of invariants.


\section{Symplectic covering of $T\times\mf t$}
\label{Ssplcov}


Recall the symplectic form $\omega$ of $G\times\mf g$, see
\eqref{Gomega}. By an abuse of notation, the pull-back of this
form to $T\times\mf t$ will also be denoted by $\omega$. Elements
of $\RR^4$ will be denoted by $(x,p)\equiv((x^1,x^2),(p_1,p_2))$.
In this section, we use the exponential map of $T$ to construct a
covering $\psi:\RR^4\to T\times\mf t$ which pulls back $\omega$ to
the natural symplectic form $\mr d p_i \wedge\mr d x^i$ of $\RR^4$
(summation convention). We choose $\psi$ to be induced by some
covering $\vp : \RR^2\to T$ by virtue of the commutative diagram
 \begin{equation}
 \label{Gdgrpsi}
 \begin{CD}
\tg\RR^2 @>\vp'>> \tg T
\\
@V g VV @VV h V
\\
\RR^4\cong\ctg\RR^2 @>\psi>> \ctg T \cong T\times\mf t
 \end{CD}
 \end{equation}
where the vertical arrows stand for the isomorphisms between the
tangent and cotangent bundles induced by the natural Riemannian
metrics $g$ on $\RR^2$ and $h$ on $T$. Recall that $h$ is given by
the restriction to $T$ of the Killing metric of $G$ induced by the
scalar product $\langle\,\cdot\,,\,\cdot\,\rangle$ on $\mf g$.
A straightforward calculation, where $\RR^2$ and $T$
may be replaced by arbitrary Riemannian manifolds, shows that
if $\vp$ is isometric then $\psi$ is symplectic.
Thus, all we have to do is to choose $\vp$ appropriately. E.g.,
we can choose $\vp$ as the composition of the isomorphism $\RR^2\to\mf t$,
mapping the canonical basis vectors $e_1$, $e_2$ to the orthonormal basis
$$
 \begin{array}{c}
\diag\big(\frac{i}{\sqrt 6}, \frac{i}{\sqrt 6}, -i\sqrt{\frac{2}{3}}\big)
 \,,~~~~~~
\diag\big(\frac{i}{\sqrt 2}, -\frac{i}{\sqrt 2}, 0\big)
 \end{array}
$$
in $\mf t$, with the exponential map $\mf t \to T$:
 \beq\label{GcovT}
\vp(x) = \diag
 \Big(
\mr e^{\mr i\left(\frac{1}{\sqrt 6} x^1 + \frac{1}{\sqrt 2}
x^2\right)}
 \,,\,
\mr e^{\mr i\left(\frac{1}{\sqrt 6} x^1 - \frac{1}{\sqrt 2}
x^2\right)}
 \,,\,
\mr e^{-\mr i\sqrt{\frac{2}{3}} x^1}
 \Big)\,.
 \eeq
 \comment{
To check that \eqref{GcovT} is indeed isometric, write tangent vectors at
$x\in\RR^2$ as $(x,X)$ with $X\in\RR^2$ and tangent vectors at $a\in T$ as $\mr
R_a'A$ with $A\in\mf t$. In these terms, the tangent map of $\vp$ reads
$$
 \begin{array}{c}
\vp'(x,X)
 =
\mr R_{\vp(x)}' \,\diag
 \left(
\mr i\big(\frac{1}{\sqrt 6} X^1 + \frac{1}{\sqrt 2} X^2\big)
 \,,\,
\mr i\big(\frac{1}{\sqrt 6} X^1 - \frac{1}{\sqrt 2} X^2\big)
 \,,\,
- \mr i\sqrt{\frac{2}{3}} X^1
 \right)
 \end{array}
$$
and the Killing metric is given by
$$
h(\mr R_a' A_1,\mr R_a' A_2) = \langle A_1,A_2\rangle = -\tr(A_1A_2)\,.
$$
Hence,
$$
h(\vp(x,X_1),\vp(x,X_2)) = X_1^1 X_2^1 + X_1^2 X_2^2 = g(X_1,X_2)\,,
$$
as asserted.}%
The corresponding covering $\psi : \RR^4 \to T\times\mf t$ is
 \beq
 \label{GcovTt}
  \begin{array}[b]{rcl}
\psi(x,p)
 & = &
 \bigg(\,
 \vp(x)\,\,,\,\,
 \diag
 \left(
\mr i\big(\frac{1}{\sqrt 6} p_1 + \frac{1}{\sqrt 2} p_2\big)
 \,,\,
\mr i\big(\frac{1}{\sqrt 6} p_1 - \frac{1}{\sqrt 2} p_2\big)
 \,,\,
- \mr i\sqrt{\frac{2}{3}} p_1
 \right)\,
 \bigg)\,.
 \comment{
 \bigg(
 \diag
 \big(
\mr e^{\mr i\left(c_1 x^1 + c_2 x^2\right)}
 \,,\,
\mr e^{\mr i(c_1 x^1 - c_2 x^2)}
 \,,\,
\mr e^{-\mr i 2c_3 x^1}
 \big)\,,
\\
 & &
 \phantom{\bigg(}\diag
 \left(
\mr i (c_1 p_1 + c_2 p_2)
 \,,\,
\mr i (c_1 p_1 - c_2 p_2)
 \,,\,
- \mr i 2 c_1 p_1
 \right)
 \bigg)\,.
 }
 \end{array}
 \eeq

\bre

Since $\psi$ is a local diffeomorphism it is a local symplectomorphism and
hence provides local Darboux coordinates on $T\times\mf t$.

\ere

Now having constructed $\psi$, we can compose it with the map
$\lambda:T\times\mf t\to\pha$, see \eqref{GdefcovT}, to obtain
 \beq\label{Gdefchi}
\chi := \lambda\circ\psi : \RR^4 \to \pha\,.
 \eeq
Let $\RR^4_k = \chi^{-1}(\pha_k)$ denote the pre-image of the
stratum $\pha_k$ under $\chi$, $k=2,1,0$. Using $\RR^4_k =
\psi^{-1}((T\times\mf t)_k)$ we find
 \beq\label{GR4R2}
 \begin{array}{c}
\RR^4_0 = \RR^2_0 \times \{0\}
 \,,~~~~~~
\RR^4_1
 =
 \left(
\bigcup_{j=1}^3 ~\bigcup_{l\in\ZZ} ~\RR^2_{(j)\,l}\times\RR^2_{(j)\,0}
 \right)
 \setminus
\RR^4_0
 \,,~~~~~~
\RR^4_2 = \RR^4  \setminus \RR^4_1\,,
 \end{array}
 \eeq
where
$$
 \begin{array}{lcl}
\RR^2_0
 & = &
\big\{~\big(\,l\,\sqrt{\frac{2}{3}}\,\pi\,,\,(l+2m)\,\sqrt 2\,
\pi\,\big)~~|~~ l,m\in\ZZ~\big\}
\\
\RR^2_{(1)\,l}
 & = &
\left\{~\big(\,y\,,\,\sqrt 3\,y + 2l\,\sqrt 2\,\pi\,\big)~~|~~ y\in\RR~\right\}
\\
\RR^2_{(2)\,l}
 & = &
\left\{~\big(\,y\,,\,-\sqrt 3\,y + 2l\,\sqrt 2\,\pi\,\big)~~|~~ y\in\RR~\right\}
\\
\RR^2_{(3)\,l}
 & = &
\left\{~\big(\,y\,,\,l\,\sqrt 2\,\pi\,\big)~~|~~ y\in\RR~\right\}\,.
 \end{array}
$$
The $\RR^2_{(j)\,l}$ are affine subspaces of $\RR^2$, intersecting each other in
the points of $\RR^2_0$, see Figure \rref{figstrataR2}.
 \begin{figure}

 \center

 \unitlength1cm

 \begin{picture}(7,5.8)
\put(0,0.5){
 \put(3.3,2.3){
 \thicklines
 \linie{0,0}{1,0}{3.3}
 \linie{0,0}{-1,0}{3.1}
 \linie{0,1}{1,0}{3.3}
 \linie{0,1}{-1,0}{3.1}
 \linie{0,2}{1,0}{3.3}
 \linie{0,2}{-1,0}{3.1}
 \linie{0,-1}{1,0}{3.3}
 \linie{0,-1}{-1,0}{3.1}
 \linie{0,-2}{1,0}{3.3}
 \linie{0,-2}{-1,0}{3.1}
 \marke{3.3,2}{cl}{\scriptscriptstyle {(1) \atop 2}}
 \marke{3.3,1}{cl}{\scriptscriptstyle {(1) \atop 1}}
 \marke{3.3,0}{cl}{\scriptscriptstyle {(1) \atop 0}}
 \marke{3.3,-1}{cl}{\scriptscriptstyle {(1) \atop -1}}
 \marke{3.3,-2}{cl}{\scriptscriptstyle {(1) \atop -2}}
 \linie{0,0}{3,5}{1.5}
 \linie{0,0}{-3,-5}{1.3}
 \linie{1.2,0}{3,5}{1.5}
 \linie{2.4,-2}{3,5}{0.7}
 \linie{1.2,0}{-3,-5}{1.3}
 \linie{2.4,0}{3,5}{0.7}
 \linie{2.4,0}{-3,-5}{1.3}
 \linie{2.3,-2.1666}{3,5}{0.8}
 \linie{-1.2,0}{3,5}{1.5}
 \linie{-1.2,0}{-3,-5}{1.3}
 \linie{-2.4,0}{3,5}{1.5}
 \linie{-2.4,0}{-3,-5}{0.7}
 \linie{-2.1,2.5}{-3,-5}{1}
 \marke{-2.1,2.5}{bc}{\scriptscriptstyle {(3) \atop 3}}
 \marke{-0.9,2.5}{bc}{\scriptscriptstyle {(3) \atop 2}}
 \marke{0.3,2.5}{bc}{\scriptscriptstyle {(3) \atop 1}}
 \marke{1.5,2.5}{bc}{\scriptscriptstyle {(3) \atop 0}}
 \marke{2.7,2.5}{bc}{\scriptscriptstyle {(3) \atop \text{-}1}}
 \linie{0,0}{3,-5}{1.3}
 \linie{0,0}{-3,5}{1.5}
 \linie{1.2,0}{3,-5}{1.3}
 \linie{1.2,0}{-3,5}{1.5}
 \linie{2.4,0}{3,-5}{0.7}
 \linie{2.4,0}{-3,5}{1.5}
 \linie{2.1,2.5}{3,-5}{1}
 \marke{-2.7,2.5}{bc}{\scriptscriptstyle {(2) \atop \text{-}1}}
 \marke{-1.5,2.5}{bc}{\scriptscriptstyle {(2) \atop 0}}
 \marke{-0.3,2.5}{bc}{\scriptscriptstyle {(2) \atop 1}}
 \marke{0.9,2.5}{bc}{\scriptscriptstyle {(2) \atop 2}}
 \marke{2.1,2.5}{bc}{\scriptscriptstyle {(2) \atop 3}}
 \linie{-1.2,0}{3,-5}{1.3}
 \linie{-1.2,0}{-3,5}{1.5}
 \linie{-2.4,0}{3,-5}{1.3}
 \linie{-2.4,0}{-3,5}{0.7}
 \linie{-2.3,-2.1666}{-3,5}{0.8}
 \multiput(-2.4,0)(1.2,0){5}{\circle*{0.17}}
 \multiput(-2.4,2)(1.2,0){5}{\circle*{0.17}}
 \multiput(-2.4,-2)(1.2,0){5}{\circle*{0.17}}
 \multiput(-3,1)(1.2,0){6}{\circle*{0.17}}
 \multiput(-3,-1)(1.2,0){6}{\circle*{0.17}}
 \multiput(0,-2)(0,2){3}{\marke{0,0.05}{bc}{\scriptscriptstyle 0    }}
 \multiput(0.6,-1)(0,2){2}{\marke{0,0.05}{bc}{\scriptscriptstyle 1}}
 \multiput(1.2,-2)(0,2){3}{\marke{0,0.05}{bc}{\scriptscriptstyle 2}}
 \multiput(1.8,-1)(0,2){2}{\marke{0,0.05}{bc}{\scriptscriptstyle 0}}
 \multiput(2.4,-2)(0,2){3}{\marke{0,0.05}{bc}{\scriptscriptstyle 1}}
 \multiput(3,-1)(0,2){2}{\marke{0,0.05}{bc}{\scriptscriptstyle 2}}
 \multiput(-0.6,-1)(0,2){2}{\marke{0,0.05}{bc}{\scriptscriptstyle 2}}
 \multiput(-1.2,-2)(0,2){3}{\marke{0,0.05}{bc}{\scriptscriptstyle 1}}
 \multiput(-1.8,-1)(0,2){2}{\marke{0,0.05}{bc}{\scriptscriptstyle 0}}
 \multiput(-2.4,-2)(0,2){3}{\marke{0,0.05}{bc}{\scriptscriptstyle 2}}
 \multiput(-3,-1)(0,2){2}{\marke{0,0.05}{bc}{\scriptscriptstyle 1}}
 }

 %
 %
 \put(0,0){\vector(1,0){7}}
 \put(0,0){\vector(0,1){4.8}}
 \marke{7,0}{cl}{\scriptstyle \frac{x^1}{\sqrt{\frac 2 3}\,\pi}}
 \marke{0,4.9}{cr}{\scriptstyle \frac{x^2}{\sqrt 2\,\pi}}
 \multiput(3.3,-0.1)(0.6,0){6}{\linie{0,0}{0,1}{0.2}}
 \multiput(3.3,-0.1)(-0.6,0){6}{\linie{0,0}{0,1}{0.2}}
 \put(3.3,-0.1){
  \marke{0,0}{tc}{\scriptstyle 0}
  \marke{0.6,0}{tc}{\scriptstyle 1}
  \marke{1.2,0}{tc}{\scriptstyle 2}
  \marke{1.8,0}{tc}{\scriptstyle 3}
  \marke{2.4,0}{tc}{\scriptstyle 4}
  \marke{3,0}{tc}{\scriptstyle 5}
  \marke{-0.6,0}{tc}{\scriptstyle \text{-}1}
  \marke{-1.2,0}{tc}{\scriptstyle \text{-}2}
  \marke{-1.8,0}{tc}{\scriptstyle \text{-}3}
  \marke{-2.4,0}{tc}{\scriptstyle \text{-}4}
  \marke{-3,0}{tc}{\scriptstyle \text{-}5}
 }
 \multiput(-0.1,2.3)(0,1){3}{\linie{0,0}{1,0}{0.2}}
 \multiput(-0.1,2.3)(0,-1){3}{\linie{0,0}{1,0}{0.2}}
 \put(-0.1,2.3){
  \marke{0,0}{cr}{\scriptstyle 0}
  \marke{0,1}{cr}{\scriptstyle 1}
  \marke{0,2}{cr}{\scriptstyle 2}
  \marke{0,-1}{cr}{\scriptstyle \text{-}1}
  \marke{0,-2}{cr}{\scriptstyle \text{-}2}
 }
 }
 \end{picture}

\caption{\label{figstrataR2} The subsets $\RR^2_0$ and
$\RR^2_{(j)\,l}$ of $\RR^2$. The elements of $\RR^2_0$
are represented by $\bullet$ and are labelled by the element of $\cfg_0$
they project to: $0,1,2$ stands for $\II$, $\exp{\mr i\frac{2}{3}\pi}\II$,
$\exp{\mr i\frac{4}{3}\pi} \II$, respectively. The affine subspaces
$\RR^2_{(j)\,l}$ are labelled by $(j)\atop l$.}

 \end{figure}
The $\RR^4_k$ are symplectic submanifolds of $\RR^4$: for $k=0$
this is trivial, for $k=2$ it is obvious. For $k=1$ it follows
from the fact that in the natural identification of $\ctg \RR^2$
with $\RR^4$ utilized here, $\RR^2_{(j)\,l}\times\RR^2_{(j)\,l}$ corresponds
to $\ctg\RR^2_{(j)\,l}$, $j=1,2,3$, $l\in\ZZ$.

By restriction, $\psi$ and $\chi$ induce maps
 \beq\label{Gedfchik}
\psi_k : \RR^4_k\to (T\times\mf t)_k
 \,,~~~~~~
\chi_k = \lambda_k\circ\psi_k : \RR^4_k \to \pha_k
 \,,~~~~~~
k = 2,1,0\,,
 \eeq
respectively.

\btm\label{TphaR4}

The map $\chi$ is Poisson, i.e., for $f,g\in C^\infty(\pha)$ there holds
$$
\chi^\ast\{f,g\}_{\!_{\scriptstyle\pha}}
 =
\frac{\partial(\chi^\ast f)}{\partial x^k}
 \,
\frac{\partial (\chi^\ast g)}{\partial p_k}
 -
\frac{\partial (\chi^\ast f)}{\partial p_k}
 \,
\frac{\partial (\chi^\ast g)}{\partial x^k}\,.
$$
The maps $\chi_k$ are local symplectomorphisms.

\etm

{\it Proof.}~ This follows from Theorem \rref{TphaTt}. In addition, for
the second assertion one has to use that the $\psi_k$ are local
symplectomorphisms. This is a consequence of the fact that $(T\times\mf t)_k$
are embedded submanifolds of  $T\times\mf t$.
 \qed


\section{Description in terms of invariants}
\label{Sivr}


In this section, we derive the invariant-theoretic description of
the reduced data of our model. Let us start with
recalling the general theory. Consider an
orthogonal representation of some Lie group $H$ on a Euclidean
space $\RR^n$. The algebra of invariant polynomials of this representation is
finitely generated \cite{Weyl:Classical}. Any finite set of generators
$\rho_1,\dots,\rho_p$ defines a map
$$
\rho =(\rho_1,\dots,\rho_p) : \RR^n/H \to\RR^p\,.
$$
This map is a homeomorphism onto its image \cite{Schwarz:Smooth} and the
image is a closed semialgebraic subset of $\RR^p$, i.e.,
it is the solution set of a logical combination of algebraic
equations and inequalities. The equations are provided by the
relations amongst the generators $\rho_i$ and the inequalities
keep track of their ranges. The set $\{\rho_1,\dots,\rho_p\}$ and
the map $\rho$ are called a Hilbert basis and a Hilbert map for
the representation, respectively. If $V\subseteq\RR^n$ is an
$H$-invariant semialgebraic subset, then $\rho$ restricts to a
homeomorphism of $V/H$ onto the image $\rho(V)\subseteq\RR^p$ and
the image is again a semialgebraic subset. The equations are now
given by the relations amongst the restricted mappings $\rho_i|_V$
and the inequalities are given by their ranges.


\subsection{Hilbert map}
\label{SSHimap}


To apply the method explained above to our model, we consider the 
realification of the representation of
$G=\SU(3)$ on $\mr M_3(\CC)\oplus\mr M_3(\CC)$ by diagonal conjugation:
 \beq\label{Gcpxrep}
a\cdot(X_1,X_2) = (aX_1a^{-1},aX_2a^{-1})
 \eeq
and set $V = J^{-1}(0)\subseteq G\times\mf g$.
Indeed, since this (complex) representation is
unitary w.r.t.\ the scalar product
 \beq\label{Gcpxscapro}
\langle (X_1,X_2),(Y_1,Y_2)\rangle
 =
\tr(X_1^\dagger Y_1) + \tr(X_2^\dagger Y_2)\,,
 \eeq
the realification, equipped with the real part of \eqref{Gcpxscapro} as
a scalar product, is orthogonal. Moreover, the subset $J^{-1}(0)\subseteq\mr
M_3(\CC)\oplus\mr M_3(\CC)$ is defined by the equations
 \beq\label{Grelslevelset}
a^\dagger a =\II
 \,,~~~~
\det(a) =1
 \,,~~~~
A^\dagger + A = 0
 \,,~~~~
aA-Aa = 0\,,
 \eeq
hence is real algebraic.

Since the invariant polynomials of the realification of a complex
representation are given by the real and imaginary parts of the
invariant polynomials of the original representation, we have to
find the generators for the latter. According to \cite{Procesi}, a
set of generators for the invariant polynomials of the
representation of $\SU(n)$ on $\mr M_n(\CC)^m$ by diagonal
conjugation is given by the trace monomials up to order $2^n-1$ in
$X_1,\dots,X_m$ and $X_1^\dagger,\dots,X_m^\dagger$. The
generators are subject to the relation
 \beq\label{GFTI}
 \sum\nolimits_{\sigma \in S_{n+1}}
 \mathrm{sgn}(\sigma)~
 \prod\nolimits_{(k_1,\ldots,k_j) \atop \text{cycle of }
 \sigma}~
 \tr(Y_{k_1}\cdots Y_{k_j})
  ~=~
 0\,,~~~~~~
 Y_1,\dots,Y_ {n+1}\in\mr M_n(\CC)\,,
 \eeq
called the fundamental trace identity (FTI).
Thus, according to the general theory, the real and imaginary parts of the
trace monomials up to order $7$ in $a,A$ and $a^\dagger,A^\dagger$, where
$(a,A)\in J^{-1}(0)$, provide a homeomorphism of $\pha$ onto a semialgebraic
subset of $\RR^p$ for some large $p$. However, for the restrictions of the trace
monomials to $J^{-1}(0)$ more relations hold than just the FTI. We can use them
to reduce the set of generators and thus to simplify the Hilbert map. They arise
from the matrix identities \eqref{Grelslevelset} and the Cayley-Hamilton
theorem which says that the characteristic polynomial $\chi_X$ of any
$X\in\mr M_n(\CC)$ obeys $\chi_X(X) = 0$. The characteristic polynomials of
$a$ and $A$ are
 \beq\label{Gcharpol}
 \begin{array}{c}
\chi_a(z) = - z^3 + \tr(a)z^2 - \ol{tr(a)} z + 1
 \,,~~~~~~
\chi_A(z) = - z^3 + \frac{1}{2} \tr(A^2) z + \frac{1}{3}
\tr(A^3)\,,
 \end{array}
 \eeq
respectively. Using \eqref{Grelslevelset}, any trace monomial can be
transformed to the form $\tr(a^kA^l)$ or its conjugate for some $k,l$. Using
\eqref{Gcharpol} it can then be rewritten as a polynomial in the monomials
$$
\tr(a)\,,~~\tr(aA)\,,~~\tr(aA^2)\,,~~\tr(A^2)\,,~~\tr(A^3)\,.
$$
We define
 \begin{align*}
c_k & := \Re(\tr(a(-iA)^k))
 \,,~~~~~~
d_k := \Im(\tr(a(-iA)^k))
 \,,~~~~~~
k=0,1,2\,,
\\
t_k & := \tr((-iA)^k)
 \,,~~~~~~
k=2,3\,.
 \end{align*}
As $iA$ is self-adjoint, $t_2$ and $t_3$ are real. Thus, we
arrive at the simplified Hilbert map
$$
\rho_\pha = (c_0,d_0,c_1,d_1,c_2,d_2,t_2,t_3) : \pha \to \RR^8\,.
$$
By embedding $G\hookrightarrow G\times \{0\} \subseteq J^{-1}(0)$,
from $\rho_\pha$ we obtain the Hilbert map for the action of $G$
on itself by inner automorphisms, i.e., for the reduced
configuration space $\cfg$:
 \beq\label{Gdefrhocfg}
\rho_\cfg = (c_0,d_0) : \cfg \to \RR^2\,.
 \eeq
Analogously, embedding $\mf g\hookrightarrow \{\II\}\times\mf g
\subseteq J^{-1}(0)$ and using that on the image of this embedding
there holds $c_2 = t_2$ and $c_1 = d_1 = d_2 = 0$, we obtain the
Hilbert map for the adjoint representation of $\SU(3)$, or the
corresponding representation of $S_3$ on $\mf t$,
$$
\rho_\Ad = (t_2,t_3) : \su(3)/\Ad \to \RR^2\,.
$$
By construction, the maps $\rho_\pha$, $\rho_\cfg$ and $\rho_\Ad$
are homeomorphisms onto their images. The images will be denoted
by $\tilde\pha$, $\tilde\cfg$ and $\tilde\rad$, respectively. The images of the
strata $\pha_k$ of $\pha$ and $\cfg_k$ of $\cfg$ will be denoted by $\tpha_k$
and $\tcfg_k$, respectively. As $\tilde\pha$, $\tilde\cfg$ and $\tilde\rad$ are
projections of a semialgebraic subset, they are semialgebraic themselves. The
reason why we consider $\trad$ is that it will be needed in the discussion of
$\tpha$.


\subsection{Reduced configuration space and quotient of adjoint representation}
\label{SStcfgtrad}


The subset $\tilde\cfg$ was discussed in \cite{configspace}. We recall the
results. A natural candidate for an inequality is given by the
discriminant $\mr D(\chi_a)$ of $\chi_a$. Indeed, as $a$ has
eigenvalues $\alpha,\beta$ and $\ol{\alpha\beta}$, where
$\alpha,\beta\in\mr U(1)$,
$$
\mr D(\chi_a)
 =
(\alpha-\beta)^2(\alpha-\ol{\alpha\beta})^2(\beta-\ol{\alpha\beta})^2
 =
- |\alpha\beta|^2|\alpha-\beta|^2
 |\alpha-\ol{\alpha\beta}|^2
 |\beta-\ol{\alpha\beta}|^2
 \leq 0\,.
$$
Define
$$
P_1(c_0(a),d_0(a)) := - \mr D(\chi_a)
 \,,~~~~~~
a\in\SU(3)\,.
$$
Expressing the discriminant in terms of the coefficients of
$\chi_a$, see \eqref{Gcharpol}, yields
$$
P_1(c_0,d_0)
 =
27 - c_0^4 - 2 c_0^2 d_0^2 - d_0^4 + 8 c_0^3 - 24 c_0 d_0^2 - 18
c_0^2 - 18 d_0^2\,.
$$
Moreover, define
$$
P_0(c_0,d_0) := 9 - c_0^2 - d_0^2\,.
$$

\btm\label{Ttcfg}

$\tcfg$ is the subset of $\RR^2$ defined by the inequality
$P_1\geq 0$. As subsets of $\tcfg$, the strata are defined by the
following equations and inequalities:
$$
\tilde\cfg_0\colon~~ P_0 = 0
  \,,~~~~~~~~
\tilde\cfg_1\colon~~ P_1 = 0 \text{ and } P_0 > 0
 \,,~~~~~~~~
\tilde\cfg_2\colon~~ P_1 > 0\,.
$$
 \comment{
 \beq\label{GstratatiX}
 \begin{array}{r|l}
\tilde\cfg_0 & P_0 = 0
\\ \hline
\tilde\cfg_1 & P_1 = 0 \text{ and } P_2 > 0
\\ \hline
\tilde\cfg_2 & P_1 > 0
 \end{array}
 \eeq
 }
\etm

{\it Proof.}~ By construction, $\tcfg$ is contained in the subset
defined by $P_1\geq 0$. The inverse inclusion was shown in
\cite{configspace}. To discuss the stratification, let $a\in\cfg$
(again identified with $\fdom$). One has $a\in\cfg_2$ iff all its
entries are distinct, i.e., iff $\mr D(\chi_a) \neq 0$. This
yields the assertion for $\tcfg_2$. On has $a\in\cfg_0$ iff all
its entries are equal. This is equivalent to $|tr(a)| = 3$, i.e.,
$c_0(a)^2 + d_0^2(a) = 9$, hence the assertion for $\tcfg_0$. Then
the assertion for $\tcfg_1$ follows.
 \qed
\bigskip

\begin{figure}
 \center
\unitlength1cm
 \begin{picture}(13.25,4.5)
 \put(1,0.5){
 \put(-1.5,0){\epsfig{file=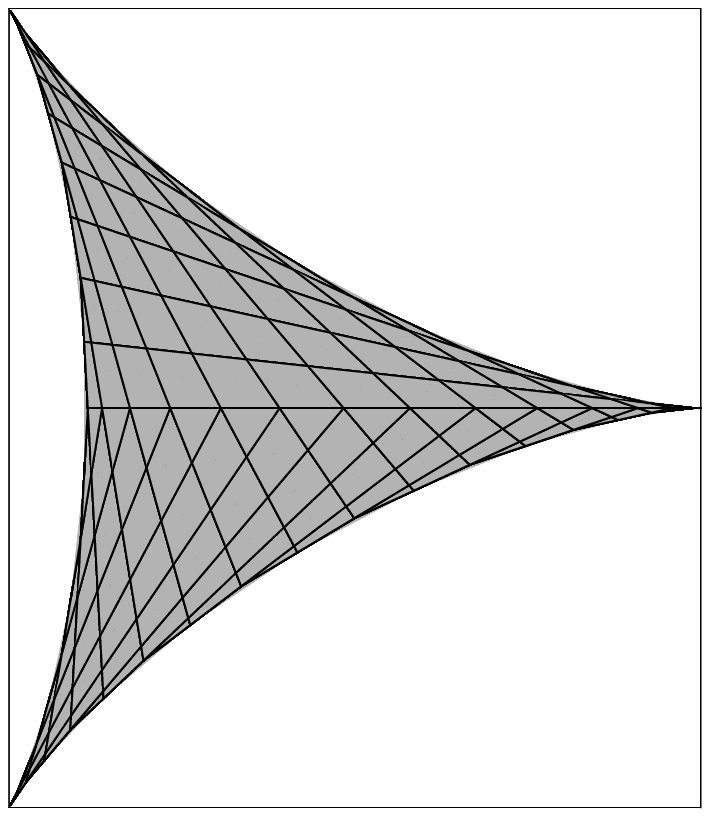,height=4cm}}
 \marke{3.55,2}{cl}{\scriptstyle (3,0)}
 \marke{0,4}{bc}{\scriptstyle (-\frac{3}{2},\frac{3}{2}\,\sqrt 3)}
 \marke{0,0}{tc}{\scriptstyle (-\frac{3}{2},-\frac{3}{2}\,\sqrt 3)}
 \marke{0,2}{cr}{\scriptstyle d_0} \marke{1.75,0}{tc}{\scriptstyle
 c_0} 
 \put(6,0){\epsfig{file=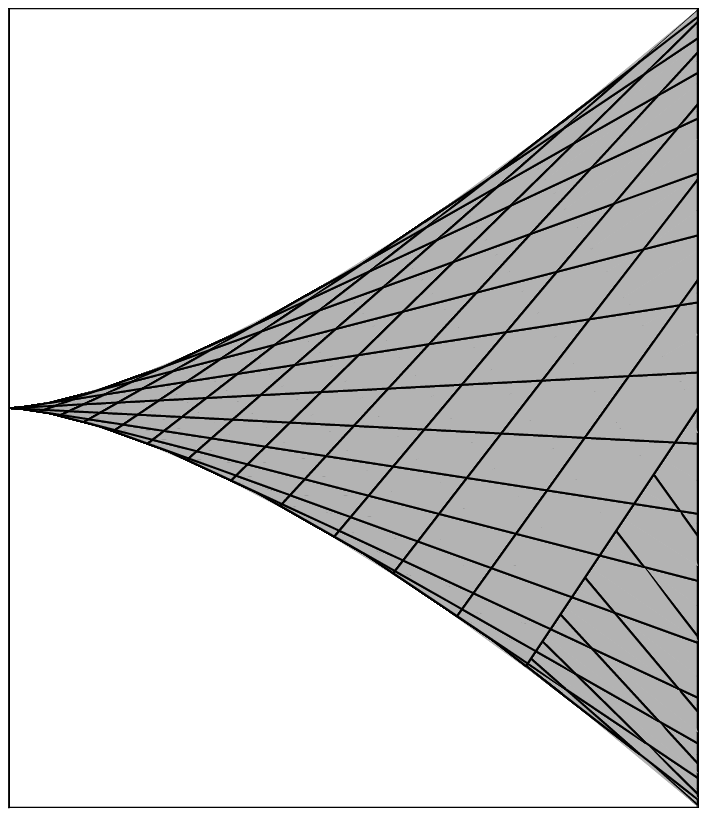,height=4cm}}
 \marke{7.5,2}{cr}{\scriptstyle (0,0)}
 \marke{7.5,3}{cr}{\scriptstyle t_3}
 \marke{9.25,0}{tc}{\scriptstyle t_2}
 }
 \end{picture}
\caption{\label{fighypocy} ~~The subsets $P_1 \geq 0$ (left) and
$P_2 \geq 0$ (right). The curve $P_1 = 0$ is a $3$-hypocycloid.
All the singular points of the curves $P_1=0$ and $P_2=0$ are
cusps.}
 \end{figure}

The curve $P_1 = 0$ is a $3$-hypocycloid in a circle of radius $3$ and $\tcfg$
is the subset of $\RR^2$ enclosed by this hypocycloid, see Figure
\rref{fighypocy}.

Next, consider $\trad$. Again, the discriminant of $\chi_A$ is a
natural candidate for an inequality: as $A$ has purely imaginary
eigenvalues, $\mr D(\chi_A) \leq 0$. Define
$$
P_2(t_2(A),t_3(A)) := - \mr D(\chi_A)
 \,~~~~~~
A\in\su(3)\,.
$$
In terms of the coefficients of $\chi_A$,
$$
P_2(t_2,t_3)
 =
\frac{1}{2} t_2^3 - 3 t_3^2\,.
$$

\ble\label{Ltrad}

$\trad$ is the subset of $\RR^2$ defined by the inequality
$P_2\geq 0$. A matrix $A\in\mf t$ has $n$ distinct entries iff the
following conditions hold:
$$
n=1\colon~~ t_2 = 0
 \,,~~~~~~~~
n=2\colon~~ P_2 = 0 \text{ and } t_2 > 0
 \,,~~~~~~~~
n=3\colon~~ P_2 > 0\,.
$$
 \comment{
  \beq\label{Gstratarad}
 \begin{array}{c|c}
n=1 & t_2 = 0
\\
n=2 & P_2 = 0 \text{ and } t_2 > 0
\\
n=3 & P_2 > 0
 \end{array}
 \eeq
 }
 \ele

{\it Proof.}~ By construction, $\trad$ is contained in the subset
of $\RR^2$ defined by $P_2\geq 0$. Conversely, for any choice of
$(t_2,t_3)\in\RR^2$ there exists $A\in\mr M_3(\CC)$ with these
values for the invariants $t_2$, $t_3$. It may be chosen as a
diagonal matrix with entries being the zeros of the polynomial
$\chi_A$, see \eqref{Gcharpol}, where the traces have to be
expressed in terms of the chosen values for $t_2$ and $t_3$. It
suffices to show that the inequality $P_2(t_2,t_3)\geq 0$ implies
$A\in\mf g = \su(3)$. Indeed, replacing $z$ by $\mr i w$ yields
$\mr i\chi_A = - w^3 + \frac{1}{2} t_2 w + \frac{1}{3} t_3$. This
polynomial has real coefficients and discriminant $-P_2(t_2,t_3)
\leq 0$. Therefore, its roots $w_1,w_2,w_3$ are real. Since it
does not contain a square term, $w_1 + w_2 + w_3 = 0$. Then
$A=\diag(\mr i w_1,\mr i w_2,\mr i w_3)\in\su(3)$.

The conditions that all entries are equal or that all
entries are distinct are obvious. The condition that $2$ entries are distinct
then follows by observing that $P_2\geq 0$ implies $t_2\geq 0$.
 \qed
\bigskip

The curve $P_2=0$ is shown in Figure \rref{fighypocy}. The
inequality $P_2\geq 0$ describes the part of the $t_2$-$t_3$-plane
to the right of this curve.


\subsection{Reduced phase space}
\label{SStpha}


Now we turn to $\tpha$. First, let us look for equations defining
$J^{-1}(0)$ inside $G\times\mf g$, i.e., reflecting the fact that
$a$ and $A$ commute. The following 2 families of functions on
$G\times\mf g$ obviously vanish on $J^{-1}(0)$:
$$
 \begin{array}{rcl}
f_k(a,A)
 & := &
2 (-i)^{k-1} \left(\tr(A^k a A a^\dagger) - \tr(A^{k+1})\right)
 \,,
\\
g_k(a,A)
 & := &
(-i)^{k-1} \left(\tr(A^k a A a) - \tr(A^{k+1}a^2)\right)
 \,, ~~~~~~ k = 1,2,\dots\,.
 \end{array}
$$
The $f_k$ and $g_k$ are polynomials on $G\times\mf g$. The $f_k$
have real coefficients and the $g_k$ have complex coefficients.
Being invariant, they can be written as polynomials in the
variables $c_k,d_k,t_k$. This way, we obtain 2 families of
equations whose common zero set contains $\rho(\pha)$. They cannot
all be independent. Indeed, for $k\geq 3$, using \eqref{Gcharpol}
one finds
$$
 \begin{array}{rcl}
f_k(a,A)
 & = &
-\frac{1}{2}\tr(A^2) f_{k-2}(a,A) + \frac{i}{3} \tr(A^3) f_{k-3}(a,A)\,,
\\
g_k(a,A)
 & = &
-\frac{1}{2}\tr(A^2) g_{k-2}(a,A) + \frac{i}{3} \tr(A^3) g_{k-3}(a,A)\,,
 \end{array}
$$
where $f_0=g_0\equiv 0$. Hence, the relevant equations are those
arising from $f_1$, $f_2$, $g_1$ and $g_2$. Taking the real and
imaginary parts -- $f_1$ and $f_2$ are already real -- we obtain the
following 6 equations:
 \begin{align}\label{Gf1}
f_1
 & =
(3 + c_0^2 + d_0^2) t_2 - 2 (c_1^2 + d_1^2) - 4(c_0 c_2 + d_0 d_2)
 = 0\,,
\\ \label{Gf2}
f_2
 & =
 \begin{array}{c}
(3 - \frac{1}{3}(c_0^2 + d_0^2)) t_3
 -
2 (c_1 c_2 + d_1 d_2) = 0\,,
 \end{array}
\\ \label{GReg1}
\Re(g_1)
 & =
c_0 c_2 - d_0 d_2 - 2 c_0 t_2 - c_1^2 + d_1^2 + 3 c_2 = 0\,,
\\ \label{GImg1}
\Im(g_1)
 & =
c_0 d_2 + d_0 c_2 + 2 d_0 t_2 - 2 c_1 d_1 - 3 d_2 = 0\,,
\\ \label{GReg2}
\Re(g_2)
 & =
 \begin{array}{c}
\frac{1}{2}( (c_0+1) c_1 - d_0 d_1) t_2
 +
(\frac{1}{3}(c_0^2 - d_0^2) - c_0) t_3 - c_1 c_2 + d_1 d_2 = 0\,,
 \end{array}
\\ \label{GImg2}
\Im(g_2)
 & =
 \begin{array}{c}
\frac{1}{2}( (c_0-1) d_1 + d_0 c_1) t_2
 +
(\frac{2}{3} c_0 d_0 + d_0) t_3 - c_1 d_2 - d_1 c_2 = 0\,.
 \end{array}
 \end{align}
These are the candidates for the equations defining $\tpha$.

Next, we look for the inequalities. Besides the two inequalities
$P_1\geq 0$ and $P_2\geq 0$ found above, which contain only pure
invariants, there is another obvious one which contains the mixed
invariants $c_2$ and $d_2$. Namely, for given $a\in T$ and
$A\in\mf t$, the entries of $a(-\mr i A)^2$ are complex numbers
whose modulus is given by the corresponding entry of $(-\mr i
A)^2$. Hence, $|\tr(a(-\mr i A)^2)| \leq \tr((-\mr i A)^2)$. In
terms of the real invariants this reads $P_3(c_2,d_2,t_2) \geq 0$,
where
$$
P_3(c_2,d_2,t_2) := t_2^2 - c_2^2 - d_2^2\,.
$$

\begin{Theorem}\label{Teqsineqs}

$\tpha$ is the subset of $\RR^8$ defined by the equations and
inequalities
 \beq\label{Geqsineqs}
f_1 = f_2 = \Re(g_1) = \Im(g_1) = \Im(g_2) = 0
 \,,~~~~~~
P_j \geq 0\,,~~ j=1,2,3\,.
 \eeq

\end{Theorem}

{\it Proof.}~ We have already checked that $\tpha$ is contained in
the subset \eqref{Geqsineqs}. In order to prove the inverse
inclusion, let there be given a point $x = (c_0,d_0,c_1,d_1,c_2,d_2,t_2,t_3)$
from the subset \eqref{Geqsineqs}. We have to show that there
exists a pair $(a,A)\in T\times\mf t$ such that $\rho_\pha(a,A) =
x$.

Due to Theorem \rref{Ttcfg} and Lemma \rref{Ltrad}, there exist
$a\in T$ and $A\in\mf t$ with $\rho_\cfg(a) = (c_0,d_0)$ and
$\rho_\Ad(A) = (t_2,t_3)$, respectively. All pairs in the orbit of
$(a,A)$ under the direct product action of $S_3\times S_3$ on
$T\times\mf t$ have the same values for the invariants
$c_0,d_0,t_2,t_3$. Hence, if in \eqref{Geqsineqs} we view
$c_0,d_0,t_2,t_3$ as fixed parameters and $c_1,d_1,c_2,d_2$ as the
variables, it suffices to show that the number $n_{\mr{sol}}$ of
distinct solutions of this system of equations and inequalities
does not exceed the number $n_{\mr{orb}}$ of orbits under the
diagonal action of $S_3$ on the $S_3\times S_3$-orbit of $(a,A)$.
This holds in particular if $n_{\mr{sol}}=1$, i.e., if the
solution is unique.

We start with separating $c_2$ and $d_2$ in the equations
$\Re(g_1)=0$ and $\Im(g_1)=0$:
 \beqa\label{Gc2}
P_0~ c_2
 & = &
(3 -c_0)\,c_1^2 - (3 - c_0)\,d_1^2 - 2 d_0\,c_1 d_1
 +
2(c_0(3 - c_0) + d_0^2)\,t_2\,,
\\ \label{Gd2}
P_0~ d_2
 & = &
d_0\,c_1^2  - d_0\,d_1^2 - 2(3 + c_0)\,c_1 d_1 + 2 d_0 (3 + 2 c_0)\,t_2\,.
 \eeqa
The inequality $P_1\geq 0$ allows for $3$ values of $c_0$, $d_0$
where the factor $P_0$ vanishes:
$$
\begin{array}{c}
(c_0,d_0) \,=\, (3,0)
 \,,\,
(-\frac{3}{2},\frac{3}{2}\sqrt 3)
 \,,\,
(-\frac{3}{2},-\frac{3}{2}\sqrt 3)\,.
 \end{array}
$$
In the first case, the combination $f_1 + 2\Re(g_1) = 0$ yields
$c_1=0$. Then \eqref{GReg1} reads $6(t_2 - c_2) + d_1^2 = 0$ and
\eqref{GImg1} reads $d_1(t_2 - c_2) = 0$. It follows $d_1=0$ and
$c_2 = t_2$. Then $P_3\geq 0$ implies $d_2 = 0$. In the other two
cases, $f_1 - 4 \Re(g_1) = 0$ implies $c_1 = d_1 = 0$. Resolving
$f_1$ for $c_2$ and inserting this into $P_3$ yields $-(\sqrt 3
t_2 \pm 2 d_2)^2 \geq 0$. Hence, $d_2 = \mp \frac{\sqrt 3}{2} t_2$
and, then, $c_2 = -\frac{1}{2} t_2$. In all $3$ cases
$n_{\mr{sol}} = 1$.

For the rest of the proof assume $P_0 \neq 0$ (due to $P_1\geq 0$
then $P_0>0$). Then $c_2$ and $d_2$ are fixed by \eqref{Gc2} and
\eqref{Gd2} and can be replaced in \eqref{Gf1} and \eqref{Gf2}:
 \beqa\label{Gf10}
2(9 + 6 c_0 - 3 c_0^2 + d_0^2)\,c_1^2
 +
2 Q_1\, d_1^2
 -
8 d_0(3+2c_0)\,c_1d_1
 -
P_1\,t_2
 & = &
0\,,
\\ \nonumber
2 (c_0-3)\,c_1^3
 +
2 d_0\,d_1^3
 +
2 d_0\,c_1^2 d_1
 +
2 (9+c_0)\,c_1 d_1^2 
 +
4 (c_0^2 - 3 c_0 - d_0^2) t_2 \, c_1
 & &
\\ \label{Gf20}
 -
(12 + 8 c_0) d_0\, t_2 \, d_1
 +
\frac{1}{3} P_0^2\, t_3
 & = &
0\,,~~~~~~
 \eeqa
where we have introduced the notation
$$
Q_1 = (3-c_0)^2 - 3 d_0^2\,.
$$
The coefficient $Q_1$ vanishes exactly for the $3$ values of
$c_0$, $d_0$ which obey $P_0 = 0$. Hence, we can solve \eqref{Gf10} for
$d_1$,
 \beq\label{Gd1}
d_1 = \frac{1}{2\,Q_1}
 \left(
(12 + 8 c_0) d_0\,c_1
 \pm
\sqrt{2\,P_1(Q_1\,t_2 - 6\,c_1^2)}
 \right)
\,.
 \eeq
If $t_2 = 0$ then $c_1 = 0$, because $d_1$ must be real, and hence
$d_1 = 0$. Due to $P_2\geq 0$, also $t_3 = 0$. Then \eqref{Gc2}
and \eqref{Gd2} imply $c_2 = d_2 = 0$. Thus, again $n_{\mr{sol}}
= 1$.

In the sequel assume $t_2\neq 0$ (due to $P_2\geq 0$ then
$t_2>0$). If $P_1 = 0$, $d_1$ is a multiple of $c_1$, hence
replacing $d_1$ in \eqref{Gf20} yields a $3$rd order polynomial
equation which has at most $3$ real solutions. I.e., $n_{\mr{sol}}
\leq 3$. On the other hand, due to Theorem \rref{Ttcfg}, $a$ has
$2$ distinct entries. Due to Lemma \rref{Ltrad}, $A$ has at least
$2$ distinct entries. Therefore, $n_{\mr{orb}} = 3$.

In what follows we assume $P_1(c_0,d_0) > 0$. Then $a$ has $3$
distinct eigenvalues.

First, consider the case $d_0 = 0$. Here, $d_1$ is a pure root and
\eqref{Gf20} contains $d_1$ only in $2$nd order. Hence, inserting
\eqref{Gd1} and discarding the global factor $\frac{(c_0 +
3)^2}{3(c_0 - 3)}$ we obtain the $3$rd order polynomial equation
 \beq\label{Gf23}
24\,c_1^3 - 3(3-c_0)^2\,t_2\,c_1 - (3 - c_0)^3\,t_3 = 0\,.
 \eeq
Since this equation has at most $3$ real roots, each of which gives rise to at most $2$
values of $d_1$ by \eqref{Gd1}, $n_{\mr{sol}}\leq 6$. It follows that in
the case $P_2 > 0$, where $A$ has $3$ distinct eigenvalues, $n_{\mr{orb}} = 6
\geq n_{\mr{sol}}$. In the case $P_2 = 0$, $A$ has $2$ distinct eigenvalues,
so that $n_{\mr{orb}} = 3$. To determine $n_{\mr{sol}}$ for this case, set
$$
x :=
 \begin{array}{c}
\sqrt[{\scriptstyle 3}]{\frac{t_3}{6}}\,.
 \end{array}
$$
Then $t_2 = 6 x^2$ and $t_3 = 6 x^3$. Substituting this in
\eqref{Gf23} and dividing by $6$ we obtain
$$
4\,c_1^3 - 3(3-c_0)^2\,x^2\,c_1 - (3 - c_0)^3\,x^3 = 0\,.
$$
Since $x\neq 0$ by assumption, the solutions of this equation are
given by $c_1 = \tilde c_1 x$, where $\tilde c_1$ are the
solutions of the same equation with $x=1$. We find $\tilde c_1 =
3-c_0$ with multiplicity $1$ and $\tilde c_1 = \frac{1}{2}(c_0 -
3)$ with multiplicity $2$. Then \eqref{Gd1} yields $d_1=0$ in the
first case and $d_1 = \pm \frac{3}{2} \sqrt{(3-c_0)(1+c_0)}\,x$ in
the second one. Thus, $n_{\mr{sol}} = 3 = n_{\mr{orb}}$.

Next, consider the case $d_0\neq 0$. We insert \eqref{Gd1} into \eqref{Gf20} and
write this equation in the form
 \beq\label{Gf2root}
\pm 3 d_0 (Q_1\,t_2 - 24\,c_1^2) \sqrt{2\,P_1(Q_1\,t_2 -
6\,c_1^2)}
 =
Q\,,
 \eeq
where $Q$ is some polynomial and we have omitted a common factor
$3 \sqrt 2 P_0^2/Q_1^3$ to avoid fractures. By
squaring \eqref{Gf2root} we obtain the $6$th order polynomial
equation in $c_1$
 \beq\label{Gf26}
1152\, c_1^6
 -
288\,Q_1\,t_2\,c_1^4
 +
96\,Q_2\,t_3\,c_1^3
 +
18\,Q_1^2\,t_2^2\,c_1^2
 -
12\,Q_3\,t_2\,t_3\,c_1
 +
2\,Q_1^3\,t_3^2 - 9\,P_1\,d_0^2\,t_2^3
 =
0\,,
 \eeq
where
$$
\begin{array}{rcl}
Q_2
 & = &
c_0^3 + 9 c_0 d_0^2 - 9 c_0^2 + 27 d_0^2 + 27 c_0 - 27\,,
\\
Q_3
 & = &
c_0^5 + 6\,c_0^3\,d_0^2 - 27\,c_0\,d_0^4 - 15\,c_0^4 - 81\,d_0^4 +
90\,c_0^3 - 162\,c_0\,d_0^2
\\
 & & - 270\,c_0^2 + 324\,d_0^2 + 405\,c_0 - 243\,,
 \end{array}
$$
and we have omitted a global factor $Q_1^3$. To a solution $c_1$
of \eqref{Gf26} for which the l.h.s.\ of \eqref{Gf2root} does not
vanish there corresponds one of the two signs in \eqref{Gf2root}
and hence by \eqref{Gd1} a unique value for $d_1$. To a solution
for which $Q_1\,t_2 - 6\,c_1^2 = 0$ there corresponds a unique
$d_1$ anyhow. To a solution for which $Q_1\,t_2 - 24\,c_1^2 = 0$
there correspond two values of $d_1$, but such a solution
necessarily has multiplicity $2$. (This phenomenon should be
interpreted the other way around: generically \eqref{Gf26} has
distinct solutions $c_1$, each with its own associated $d_1$. When
$2$ of the solutions happen to coincide, the associated values of
$d_1$ seem to emerge from the same $c_1$.) From these observations
we conclude that $n_{\mr{sol}} \leq 6$, so that for $P_2 > 0$ we
have $n_{\mr{orb}} = 6 \geq n_{\mr{sol}}$.

It remains to consider the case $P_2 = 0$, where $n_{\mr{orb}} =
3$. As before, we replace $t_2 = 6x^2$ and $t_3 = 6 x^3$ in
\eqref{Gf26} and argue that the solutions of the resulting
equation are given by $c_1 = \tilde c_1 x$, where $\tilde c_1$ are
the solutions of this equation with $x$ set to $1$. The latter
equation turns out to be the square of
 \beq\label{Gf2x1}
4\,\tilde c_1^3 - 3 Q_1\,\tilde c_1 + Q_2 = 0\,,
 \eeq
hence it has at most $3$ distinct real solutions $\tilde c_1$. We
claim that for none of the corresponding solutions $c_1 = \tilde
c_1 x$ the factor $Q_1 t_2 - 24 c_1^2 = 6 x \left(Q_1 - 4 \tilde
c_1^2\right)$ in \eqref{Gf2root} vanishes. Assume, on the
contrary, $Q_1 - 4 \tilde c_1^2 = 0$. Inserting $\tilde c_1 =
\pm\sqrt{Q_1}$ into \eqref{Gf2x1} and separating the terms with
the root yields $\pm Q_1\,\sqrt{Q_1} = Q_2$. Taking the square we
obtain $27\,d_0^2 \,P_1 = 0$, in contradiction to the assumptions
$d_0\neq 0$ and $P_1\neq 0$. It follows that to each $c_1$ there
corresponds a unique value for $d_1$. Thus, $n_{\mr{sol}} = 3 =
n_{\mr{orb}}$.

This completes the proof of Theorem \rref{Teqsineqs}.
 \qed

\bre

As a by-product of the proof we have seen that the $6$ invariants
$c_0$, $d_0$, $c_1$, $d_1$, $t_2$, $t_3$ are sufficient to
separate the points of $\pha$. Hence, they define a homeomorphism
of $\pha$ onto the projection of $\tpha$ to $\RR^6$. (Outside some
'momentum cutoff' $\| A\| \leq k$ the homeomorphism property is
obvious and inside one uses that a bijection of a compact space
onto a Hausdorff space is a homeomorphism.) The invariants $c_2$,
$d_2$ cannot be expressed as polynomials in the other invariants,
though. However, according to \eqref{Gc2} and \eqref{Gd2} and the
subsequent discussion, on $\tpha$ they can be expressed as
continuous functions in the other invariants. For $(c_0,d_0) \neq
(3,0), (-\frac{3}{2},\pm\frac{3}{2}\sqrt 3)$,
 \beqa\label{Gc2res}
c_2
 & = &
P_0^{-1}\,\Big(\,(3 -c_0)\,c_1^2 - (3 - c_0)\,d_1^2 - 2 d_0\,c_1
d_1
 +
2(c_0(3 - c_0) + d_0^2)\,t_2\,\Big) \,,
\\ \label{Gd2res}
d_2
 & = &
P_0^{-1}\, \Big(\,d_0\,c_1^2  - d_0\,d_1^2 - 2(3 + c_0)\,c_1 d_1 +
2 d_0 (3 + 2 c_0)\,t_2\,\Big)\,,
 \eeqa
whereas for $(c_0,d_0) = (3,0), (-\frac{3}{2},\pm\frac{3}{2}\sqrt 3)$, in the
respective order,
 \beq\label{Gc2d2vertices}
\begin{array}{c}
(c_2,d_2) = (3\,t_2,0)\,,~(-\frac{1}{2}\,t_2,\mp\frac{\sqrt 3}{2}\,t_2)\,.
\end{array}
 \eeq
One can extend $c_2$ and $d_2$ to rational functions on $\RR^6$ by
means of the expressions on the r.h.s.\ of \eqref{Gc2res} and
\eqref{Gd2res}. Then the values \eqref{Gc2d2vertices} have to be
understood as the limits when $(c_0,d_0)\to (3,0),
(-\frac{3}{2},\pm\frac{3}{2}\sqrt 3)$ along $\tpha$.

\ere

\begin{figure}
 \center
\unitlength1cm
 \begin{picture}(14.5,5.6)
 \put(0,-0.3){
\put(-1.7,1){\epsfig{file=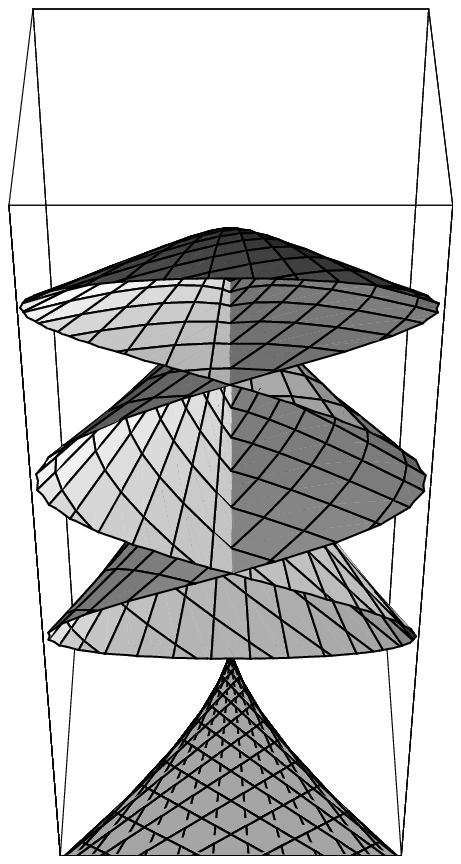,width=4.2cm}}
\marke{2.3,1.05}{cl}{\scriptstyle t_2}
\marke{0.1,2.25}{cr}{\scriptstyle t_3}
\marke{0,4.6}{cr}{\scriptstyle c_1\,}
\marke{1.25,0.3}{bc}{\scriptstyle (c_0,d_0) \in \tilde\cfg_2}
\marke{1.25,5}{bc}{\mbox{\footnotesize (a)}}
\put(1.3,1){\epsfig{file=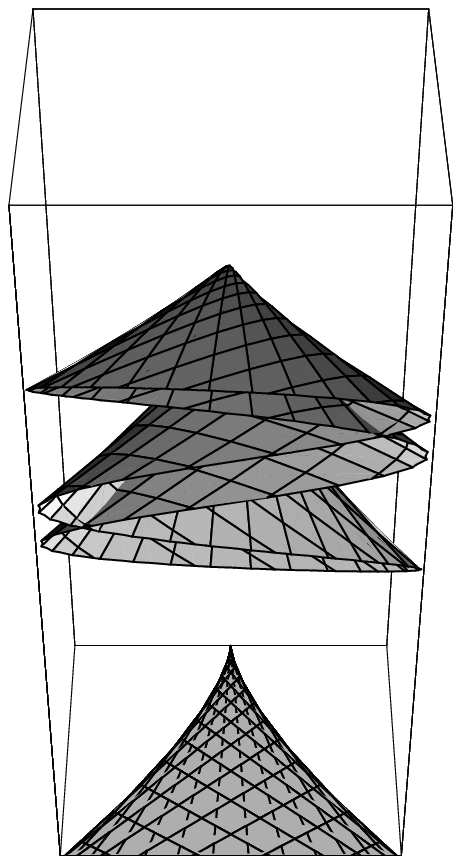,width=4.2cm}}
\marke{4.25,0.3}{bc}{\scriptstyle (c_0,d_0) \to \tilde\cfg_1}
\marke{4.25,5}{bc}{\mbox{\footnotesize (b)}}
\put(4.3,1){\epsfig{file=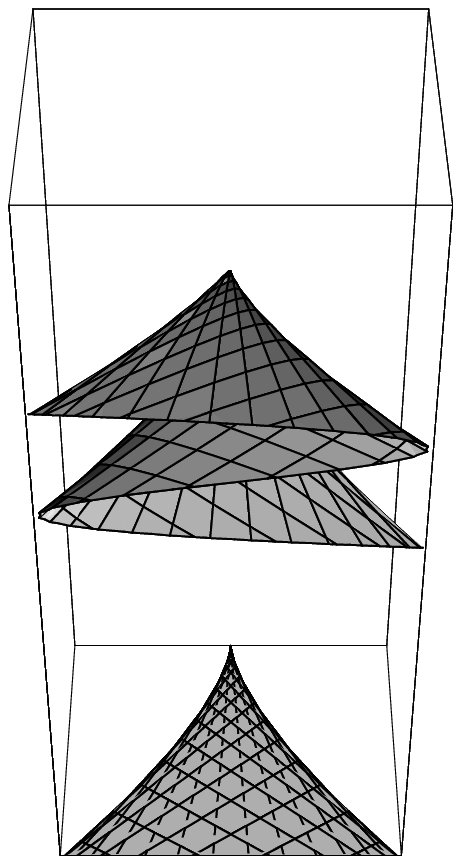,width=4.2cm}}
\marke{7.25,0.3}{bc}{\scriptstyle (c_0,d_0) \in \tilde\cfg_1}
\marke{7.25,5}{bc}{\mbox{\footnotesize (c)}}
\put(7.3,1){\epsfig{file=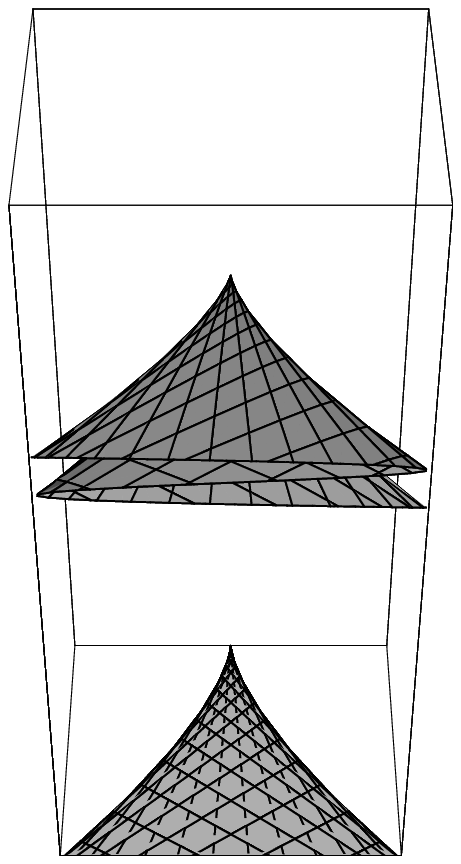,width=4.2cm}}
\marke{10.25,0.3}{bc}{\scriptstyle (c_0,d_0) \to \tilde\cfg_0}
\marke{10.25,5}{bc}{\mbox{\footnotesize (d)}}
\put(10.3,1){\epsfig{file=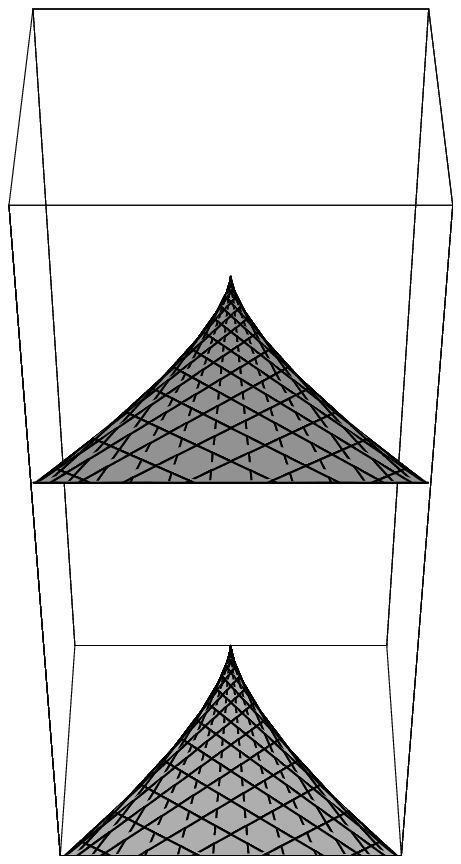,width=4.2cm}}
\marke{13.25,0.3}{bc}{\scriptstyle (c_0,d_0) \in \tilde\cfg_0}
\marke{13.25,5}{bc}{\mbox{\footnotesize (e)}}
 }
 \end{picture}
\caption{\label{Figfibre} Projection of the fibres
$\tilde\pi^{-1}(c_0,d_0)$ to the $t_2$-$t_3$-$c_1$ plane (top) and the
$t_2$-$t_3$ plane (bottom).}
 \end{figure}

On the level of the semialgebraic sets, the projection 
$\tilde\pi : \tilde\pha\to \tilde\cfg$ is just
given by the natural projection to the $c_0$-$d_0$ plane. Figure
\rref{Figfibre} shows the projections of the fibres
$\tilde\pi^{-1}(c_0,d_0)$ to the $3$-dimensional subspace spanned
by the coordinates $t_2$, $t_3$ and $c_1$ for $5$ different points
$(c_0,d_0)\in\tilde\cfg$. In addition, the projection of the
fibres to the $t_2$-$t_3$ plane, which just coincides with
$\trad$, is shown, too. The figures where drawn using a
parametrization of the invariants, induced by a parametrization of
the matrices $a$ and $A$.

For $(c_0,d_0)$ belonging to the stratum $\tilde\cfg_2$ the fibre
$\tilde\pi^{-1}(c_0,d_0)$ is a full $2$-plane, folded 3 times over
the curve $P_2 = 0$ (Figures (a) and (b)). The self-intersections
present in these figures are remnants of the projection to the
$t_2$-$t_3$-$c_1$ hyperplane. In fact, they correspond to
solutions $c_1$ of \eqref{Gf26} of multiplicity $2$ where the
factor $Q_1\,t_2-24\,c_1^2$ in \eqref{Gf2root} vanishes, so that
both values of $d_1$ in \eqref{Gd1} are allowed. Since the latter
are distinct (unless $t_2=0$), the fictitious self-intersection of
the fibre is not present in the full $\RR^8$.

When $(c_0,d_0)$ approaches the stratum $\tilde\cfg_1$, i.e., the
curve $P_1=0$, the two halves of the plane come closer (Figure
(b)). For $(c_0,d_0)\in\tilde\cfg_1$ they meet each other and thus
make the fibre a half-plane with a double fold (Figure (c)). When
moving $(c_0,d_0)$ further along $\tilde\cfg_1$ towards one of the
points of the stratum $\tilde\cfg_0$ the three layers of this
half-plane approach each other (Figure (d)) to finally merge to a
'sixth-plane' cone for $(c_0,d_0)\in\tilde\cfg_0$ (Figure (e)).
This illustrates the abstract description of the fibres in Section
\rref{SSredcotabun}.


\subsection{Stratification}
\label{SSstrativr}


We determine the equations and inequalities defining the
strata of $\tilde\pha$. We will make use of the discriminant of
the characteristic polynomial $\chi_{aA}$. Define
$$
P_4(c_0(a,A),\dots,t_3(a,A)) :=  \Re(\mr D(\chi_{aA}))\,.
$$
Using \eqref{GFTI} and \eqref{Gcharpol} one finds
$$
 \begin{array}{c}
\chi_{aA}(z)
 =
 -
z^3
 +
\tr(aA)\,z^2
 +
\left(\frac{1}{2}\,\tr(A^2)\,\ol{\tr(a)} - \ol{\tr(aA^2})\right)\,z
 +
\frac{1}{3} \tr(A^3)\,.
 \end{array}
$$
It follows
$$
 \begin{array}{rcl}
P_4
 & = &
c_1^2 c_2^2
- c_1^2 d_2^2
+ 4 \, c_1 d_1 c_2 d_2
+ d_1^2 d_2^2
- \frac{4}{3} t_3 \, c_1^3
- c_0 t_2 \, c_1^2 c_2
+ d_0 t_2 \, c_1^2 d_2
+ 4 t_3 \, c_1 d_1^2
\\ & &
- 2 d_0 t_2 \, c_1 d_1 c_2
- 2 c_0 t_2 \, c_1 d_1 d_2
+ c_0 t_2\,d_1^2 c_2
- d_0 t_2\,d_1^2 d_2
- 4 c_2^3
\\ & &
+ 12 \, c_2 d_2^2
+ \frac{1}{4} (c_0^2 - d_0^2) t_2^2 \, c_1^2
+ c_0 d_0 t_2^2 \, c_1 d_1
+ 6 t_3 \, c_1 c_2
- \frac{1}{4} (c_0^2 - d_0^2) t_2^2 \, d_1^2
\\ & &
+ 6 t_3 \, d_1 d_2
+ 6 c_0 t_2 \, c_2^2
- 12 d_0 t_2 \, c_2 d_2
- 6 c_0 t_2 \, d_2^2
- 3 c_0 t_2 t_3 \, c_1
\\ & &
- 3 d_0 t_2 t_3 \, d_1
+ 3 (d_0^2 - c_0^2) t_2^2 \, c_2
+ 6 c_0 d_0 t_2^2 \, d_2
+ \frac{1}{2} c_0 (c_0^2 - 3 d_0^2) t_2^3 - 3 t_3^2
 \end{array}
$$

\btm\label{Tstrfic}

As subsets of $\tilde\pha$, the strata $\tpha_k$ are defined by the following
equations and inequalities:
$$
 \begin{array}{rcl}
\tilde\pha_0 & : & P_0 = 0 \text{ and } t_2 = 0
\\
\tilde\pha_1 & : & P_1 = P_2 = P_4 = 0 \text{ and } (P_0 > 0
\text{ or } t_2 > 0)
\\
\tilde\pha_2 & : & P_1 > 0 \text{ or } P_2 > 0 \text{ or } P_4
\neq 0
 \end{array}
$$

\etm

{\it Proof.}~ Let $(a,A)\in T\times\mf t$ be given.

The pair $(a,A)$ is invariant under the full $S_3$-action iff so
are $a$ and $A$ individually. According to Theorem \rref{Ttcfg}
and Lemma \rref{Ltrad}, this holds iff $P_0 = 0$ and $t_2 = 0$,
respectively. Next, assume that $(a,A)$ has nontrivial stabilizer.
Then there are $2$ entries which coincide for $a$ and $A$
simultaneously. Then $aA$ has a degenerate eigenvalue. It follows
$\mr D(\chi_{aA}) = 0$ and, hence, $P_4 = 0$. Conversely, assume
$P_1 = P_2 = P_4 = 0$. Then $a$ and $A$ both have coinciding
entries. Up to $S_3$-action we can assume
$a=\diag(\alpha,\alpha,\ol\alpha^2)$, $\alpha\in\mr U(1)$. Then
$A$ can be
 \beq\label{GcasesA}
\diag(\mr ix,\mr ix,-2\mr ix)
 \,,~~
\diag(\mr ix,-2\mr ix,\mr ix)
 ~\text{ or }~
\diag(-2\mr ix,\mr ix,\mr ix)
 \,,~~~~x\in\RR\,.
 \eeq
If $x = 0$ or $\alpha^3 = 1$ then in all $3$ cases $(a,A)$ has
nontrivial stabilizer. Hence, assume $x\neq 0$ and $\alpha^3 \neq
1$. In the second and the third case,
$$
\mr D(\chi_{aA})
 =
(\alpha x + 2 \alpha x)^2 (\alpha x - \ol\alpha^2 x)^2 (2 \alpha x
+ \ol\alpha^2 x)^2
 =
9 x^6 (2 \alpha^3 - \ol\alpha^3 - 1)^2
$$
Taking the real part and replacing $\Im(\alpha^3)^2 = 1 -
\Re(\alpha^3)^2$ yields
$$
P_4 = 72 x^6 (\Re(\alpha^3) - 1)^2 = 0\,.
$$
Hence, $x=0$ or $\alpha^3 = 1$, in contradiction to the
assumption. Therefore, $A = \diag(\mr ix,\mr ix,-2\mr ix)$ and
hence $(a,A)$ has nontrivial stabilizer. This yields the equations
for $\tilde\pha_1$. The inequalities for $\tilde\pha_1$ and
$\tilde\pha_2$ are obvious.
 \qed
\bigskip


\subsection{Poisson structure}
\label{SSPoistrucivr}


The brackets of the generating invariants $c_0,\dots,t_3$, taken in the Poisson
algebra $C^\infty(\pha)$, define a Poisson structure on $\RR^8$ by
 \beq\label{GPoistrucR8}
\{f,g\}
 :=
 \sum\nolimits_{i,j=1}^8
\frac{\partial f}{\partial x_i} \frac{\partial g}{\partial x_j}
\{x_i,x_j\}\,,
 \eeq
where $(x_1,\dots,x_8) = (c_0,\dots,t_3)$. This Poisson structure rules the
dynamics on $\tpha$, see the brief remark in Section \rref{Sdynamics}.
The Poisson brackets in $C^\infty(\pha)$ are defined by
$$
\{f,g\} = \omega(X_f,X_g)
 \,,~~~~~~
f,g\in C^\infty(G\times\mf g)\,,
$$
where the symplectic form $\omega$ is given by \eqref{Gomega} and $X_f$, $X_g$
are the Hamiltonian vector fields associated with $f$ and $g$, respectively.
They are defined pointwise by
 \beq\label{GdefHaVF}
\omega_{(a,A)} (X_f,X) = - X(f)
 \,,~~~~~~
 \eeq
for all $X\in\tg_{(a,A)}(G\times\mf g)$ and $(a,A)\in G\times\mf g$. Here
$X(f)$ is the directional derivative of $f$ along $X$. As in Subsection
\rref{SSredphaspa} we write the tangent vectors in the form
$$
(X_f)_{(a,A)} = (\mr R_a' B_f,(A,C_f))
 \,,~~~~~~
X = (\mr R_a' B,(A,C))
$$
with $B_f,C_f,B,C \in \mf g$. Although it is not indicated by the notation,
$B_f$ and $C_f$ depend on $a$ and $A$, i.e., they are $\mf
g$-valued functions on $G\times\mf g$. Using \eqref{Gomega} and the invariance
of the scalar product $\langle\,\cdot\,,\,\cdot\,\rangle$ to rewrite the
l.h.s.\ of \eqref{GdefHaVF}, and using the curve $(\exp(tB)\,a\,,\,A+tC)$ to
represent $X$, \eqref{GdefHaVF} becomes
 \beq\label{GHaVF}
\langle B_f,C\rangle + \langle [B_f,A] - C_f,B\rangle
 =
- \left. \frac{\mr d}{\mr d t} \right|_{t=0} f((\exp(tB)\,a\,,\,A+tC))
 \,,~~~~~~
\forall~B,C\in\mf g\,.
 \eeq
Putting $B=0$ yields $B_f$, then putting $C=0$ and replacing $B_f$ in the
commutator yields $C_f$. Having found the Hamiltonian vector fields associated
with the invariants this way, the Poisson brackets are then given pointwise by
 \beq\label{GfmlPoibra}
\{ f, g\}((a,A))
 =
\langle B_f,C_g \rangle - \langle C_f,B_g \rangle - \langle A,[B_f,B_g]\rangle
 \eeq
Since it suffices to compute the brackets on the level set $J^{-1}(0)$, we may
always assume $(a,A)\in J^{-1}(0)$. This simplifies the computations
considerably. In particular, the commutators in \eqref{GHaVF} and
\eqref{GfmlPoibra} happen to vanish.

Let us illustrate the calculation by the bracket $\{c_1,d_1\}$.
For $c_1$ and $d_1$, \eqref{GHaVF} reads
$$
 \begin{array}{rcl}
\langle B_{c_1},C\rangle + \langle [B_{c_1},A] - C_{c_1},B\rangle
 & = &
- \Im\langle a,C \rangle - \Im\langle aA,B \rangle
 \\
\langle B_{d_1},C\rangle + \langle [B_{d_1},A] - C_{d_1},B\rangle
 & = &
- \Re\langle a,C \rangle - \Re\langle aA,B \rangle
 \end{array}
$$
To express the r.h.s.\ in terms of scalar products of $B$ and $C$ with
elements of $\mf g$, let $\Pi_+$ and $\Pi_-$ denote the projections of
$\mr M_3(\CC)$ onto the traceless Hermitian and traceless anti-Hermitian
matrices, respectively. I.e.,
$$
 \begin{array}[b]{c}
\Pi_\pm (D) = \frac{1}{2} \big(D \pm D^\dagger\big) - \frac{1}{6}\big(\tr(D) \pm
\ol{\tr(D)}\big)
 \,,~~~~~~
D\in\mr M_3(\CC)\,.
 \end{array}
$$
Both $\Pi_-$ and $\mr i\Pi_+$ map $\mr M_3(\CC)$ to $\mf g$ and
for any $D\in\mr M_3(\CC)$ and $B\in\mf g$ one has
 \beq\label{ReImprj}
\Re\langle D,B \rangle = \langle  \Pi_-(D),B\rangle
 \,,~~~~~~
\Im\langle D,B \rangle = \langle  \mr i\Pi_+(D),B\rangle\,.
 \eeq
This way, we obtain the Hamiltonian vector fields of the invariants:
$$
\renewcommand{\arraystretch}{1.15}
\arraycolsep3pt
 \begin{array}[b]{rclcrcl}
B_{c_0} & = & 0
 \,, & ~ &
C_{c_0} & = & - \Pi_-(a)
 =
- \frac{1}{2} (a - a^\dagger) + \frac{\mr i}{3} d_0\,,
\\
B_{d_0} & = & 0
 \,, & ~ &
C_{d_0} & = & \mr i\Pi_+(a)
 =
\frac{\mr i}{2} (a + a^\dagger) - \frac{\mr i}{3} c_0\,,
\\
B_{c_1} & = & - \mr i \Pi_+(a)
 =
-\frac{\mr i}{2}(a+a^\dagger) + \frac{\mr i}{3} c_0
 \,, & ~ &
C_{c_1} & = & \mr i \Pi_+(aA)
 =
\frac{\mr i}{2}(a - a^\dagger)A + \frac{\mr i}{3} d_1\,,
\\
B_{d_1} & = & - \Pi_-(a)
 =
- \frac{1}{2}(a - a^\dagger) + \frac{\mr i}{3} d_0
 \,, & &
C_{d_1} & = & \Pi_-(aA)
 =
\frac{1}{2}(a + a^\dagger)A - \frac{\mr i}{3} c_1\,,
\\
B_{c_2} & = & - 2 \Pi_-(aA)
 =
- (a + a^\dagger)A + \frac{2\mr i}{3} c_1
 \,, & ~ &
C_{c_2} & = & \Pi_- (aA^2)
 =
\frac{1}{2}(a - a^\dagger)A^2 + \frac{\mr i}{3} d_2\,,
\\
B_{d_2} & = & 2\mr i\Pi_+(aA)
 =
\mr i (a - a^\dagger)A + \frac{2\mr i}{3} d_1
 \,, & ~ &
C_{d_2} & = & -\mr i\Pi_+ (aA^2)
 =
- \frac{\mr i}{2}(a + a^\dagger)A^2 - \frac{\mr i}{3} c_2\,,
\\
B_{t_2} & = & -2 A
 \,, & ~ &
C_{t_2} & = & 0\,,
\\
B_{t_3} & = & 3\mr i \Pi_+(A^2)
 =
3\mr i A^2 + \mr i t_2
 \,, & ~ &
C_{t_3} & = & 0\,.
 \end{array}
$$
There hold the relations $B_{c_1} = -C_{d_0}$, $B_{d_1} = -C_{c_0}$, $B_{c_2} =
- 2 C_{d_1}$, $B_{d_2} = 2 C_{c_1}$. According to \eqref{GfmlPoibra}, e.g.,
$$
 \begin{array}[b]{rcl}
\{c_1,d_1\}
 & = &
\langle B_{c_1},C_{d_1}\rangle - \langle C_{c_1},B_{d_1}\rangle
 =
\langle - \mr i \Pi_+(a),C_{d_1}\rangle - \langle \mr i
\Pi_+(aA),B_{d_1}\rangle
\\
 & = &
-\Im\langle a,C_{d_1} \rangle - \Im\langle aA,B_{d_1} \rangle\,.
 \end{array}
$$
By replacing $C_{d_1}$ and $B_{c_1}$ using the above explicit expressions and
rewriting the resulting scalar products in terms of the invariants
$c_0,\dots,t_3$ we finally arrive at the desired Poisson brackets:
 \beq\label{GPoibraivr}
\renewcommand{\arraystretch}{1.15}
 \begin{array}[b]{rclcrcl}
\{c_0,d_0\} & = & 0
 \,, & ~~ &
\{c_1,d_1\} & = & \frac{1}{3} (c_0 c_1 + d_0 d_1)
\\
\{t_2,t_3\} & = & 0
 \,, & ~~ &
\{c_2,d_2\} & = & - 2 t_3 + \frac{2}{3} (c_1 c_2 + d_1 d_2)
\\[0.3cm]
\{c_0,c_1\} & = & -\frac{2}{3} c_0 d_0 - d_0
 \,, & ~~ &
\{d_0,d_1\} & = & \frac{2}{3} c_0 d_0 + d_0
\\
\{c_0,d_1\} & = & \frac{1}{2} c_0^2 - \frac{1}{6} d_0^2 - c_0 - \frac{3}{2}
 \,, & ~~ &
\{d_0,c_1\} & = & \frac{1}{6} c_0^2 - \frac{1}{2} d_0^2 - c_0 + \frac{3}{2}
\\
\{c_0,c_2\} & = & - c_0 d_1 - \frac{1}{3} d_0 c_1 + d_1
 \,, & ~~ &
\{d_0,d_2\} & = & \frac{1}{3} c_0 d_1 + d_0 c_1 - d_1
\\
\{c_0,d_2\} & = & c_0 c_1 - \frac{1}{3} d_0 d_1 + c_1
 \,, & ~~ &
\{d_0,c_2\} & = & \frac{1}{3} c_0 c_1 - d_0 d_1 + c_1
\\[0.3cm]
\{c_1,c_2\}
 & = &
\multicolumn{5}{l}{- \frac{5}{6} c_0 d_2 - \frac{1}{2} d_0 c_2 - \frac{1}{2} d_0 t_2
+ \frac{1}{2} d_2 + \frac{2}{3} c_1 d_1}
\\
\{c_1,d_2\}
 & = &
\multicolumn{5}{l}{\frac{5}{6} c_0 c_2 - \frac{1}{2} d_0 d_2 - \frac{1}{2} c_0 t_2 -
\frac{3}{2} t_2 + \frac{1}{2} c_2 + \frac{2}{3} d_1^2}
\\
\{d_1,c_2\}
 & = &
\multicolumn{5}{l}{\frac{1}{2} c_0 c_2 - \frac{5}{6} d_0 d_2 - \frac{1}{2} c_0 t_2 +
\frac{1}{2} c_2 + \frac{3}{2} t_2 - \frac{2}{3} c_1^2}
\\
\{d_1,d_2\}
 & = &
\multicolumn{5}{l}{\frac{1}{2} c_0 d_2 + \frac{5}{6} d_0 c_2 + \frac{1}{2} d_0 t_2 -
\frac{1}{2} d_2 - \frac{2}{3} c_1 d_1}
\\[0.3cm]
\{c_0,t_2\} & = &  - 2 d_1
 \,, & ~~ &
\{d_0,t_2\} & = &  2 c_1
\\
\{c_1,t_2\} & = &  -2 d_2
 \,, & ~~ &
\{d_1,t_2\} & = &  2 c_2
\\
\{c_2,t_2\} & = &  - t_2 d_1 - \frac{2}{3} t_3 d_0
 \,, & ~~ &
\{d_2,t_2\} & = &  t_2 c_1 + \frac{2}{3} t_3 c_0
\\[0.3cm]
\{c_0,t_3\} & = &  t_2 d_0 - 3 d_2
 \,, & ~~ &
\{d_0,t_3\} & = &   - t_2 c_0 + 3 c_2
\\
\{c_1,t_3\} & = &  -\frac{1}{2} t_2 d_1 - t_3 d_0
 \,, & ~~ &
\{d_1,t_3\} & = &  \frac{1}{2} t_2 c_1 + t_3 c_0
\\
\{c_2,t_3\} & = &  - \frac{1}{2} t_2 d_2 - t_3 d_1
 \,, & ~~ &
\{d_2,t_3\} & = &  \frac{1}{2} t_2 c_2 + t_3 c_1
 \end{array}
 \eeq

\bre

Another description of the reduced phase space in terms of
invariants can be constructed as follows \cite{Hue2,Hue3}. The polar map
$(a,A)\mapsto a \exp(-iA)$ yields a diffeomorphism of $T\times\mf
t$ onto the complexification $T^\CC$, which is isomorphic to the
direct product of two copies of the group of nonzero complex
numbers. This diffeomorphism passes to an isomorphism of
stratified symplectic space from $\pha$ onto $T^\CC/S_3$. The real
invariants for the latter quotient are the elementary bisymmetric
functions on $T^\CC$, obtained from the elementary symmetric
functions by bilinearization w.r.t.\ the holomorphic coordinates
and their complex conjugates. This description is the starting
point for stratified K\"ahler quantization in \cite{Hue1,Hue2}. It also
has the great advantage that it directly generalizes to $\SU(n)$
and further to an arbitrary compact Lie group. For classical
dynamics, however, it has the drawback that the kinetic energy is
not polynomial in the generating invariants.

\ere


\section{Towards classical dynamics (an outlook)}
\label{Sdynamics}


In this final section, we make some general remarks on the dynamics on $\pha$
and $\cfg$. A detailed study will be carried out in a subsequent paper.

\begin{figure}

 \centering

\unitlength1cm

 \begin{picture}(6.35085,5.5)
 \put(0,0.5){
 \put(-2.47,-0.072){\epsfig{file=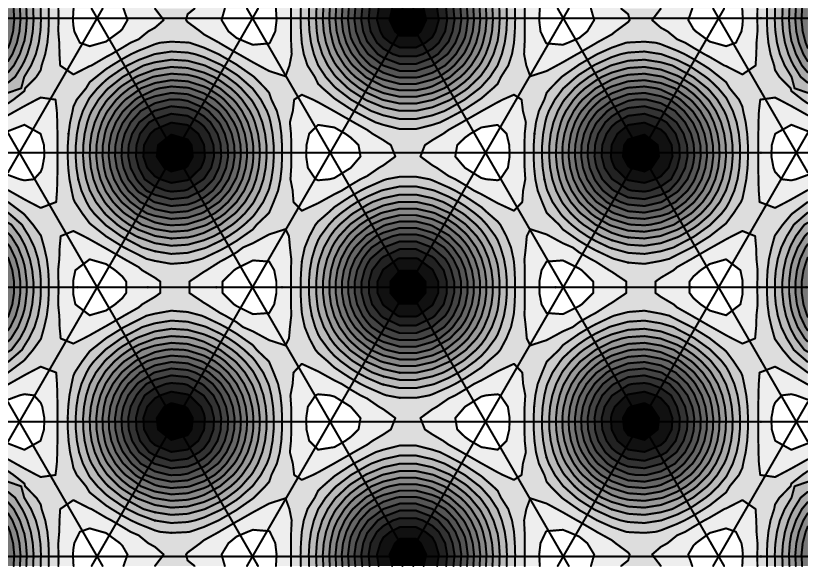,height=4.465cm}}
 %
 %
 \put(0,0){\vector(1,0){6.60085}}
 \put(0,0){\vector(0,1){4.65}}
 \marke{6.60085,0}{cl}{\scriptstyle \frac{x^1}{\sqrt{\frac 2 3}\,\pi}}
 \marke{0,4.75}{cr}{\scriptstyle \frac{x^2}{\sqrt 2\,\pi}}
 \multiput(3.164,-0.1)(0.6118,0){6}{\linie{0,0}{0,1}{0.2}}
 \multiput(3.164,-0.1)(-0.6118,0){6}{\linie{0,0}{0,1}{0.2}}
 \put(3.164,-0.1){
  \marke{0,0}{tc}{\scriptstyle 0}
  \marke{0.6118,0}{tc}{\scriptstyle 1}
  \marke{1.2236,0}{tc}{\scriptstyle 2}
  \marke{1.8354,0}{tc}{\scriptstyle 3}
  \marke{2.4472,0}{tc}{\scriptstyle 4}
  \marke{3.059,0}{tc}{\scriptstyle 5}
  \marke{-0.6118,0}{tc}{\scriptstyle \text{-}1}
  \marke{-1.2236,0}{tc}{\scriptstyle \text{-}2}
  \marke{-1.8354,0}{tc}{\scriptstyle \text{-}3}
  \marke{-2.4472,0}{tc}{\scriptstyle \text{-}4}
  \marke{-3.059,0}{tc}{\scriptstyle \text{-}5}
 }
 \multiput(-0.1,2.2)(0,1.0545){3}{\linie{0,0}{1,0}{0.2}}
 \multiput(-0.1,2.2)(0,-1.0545){3}{\linie{0,0}{1,0}{0.2}}
 \put(-0.1,2.2){
  \marke{0,0}{cr}{\scriptstyle 0}
  \marke{0,1.0545}{cr}{\scriptstyle 1}
  \marke{0,2.109}{cr}{\scriptstyle 2}
  \marke{0,-1.0545}{cr}{\scriptstyle \text{-}1}
  \marke{0,-2.109}{cr}{\scriptstyle \text{-}2}
 }
 }
 \end{picture} 

 \caption{\label{figpotential} Level diagram of the potential of the
 Hamiltonian \eqref{GHR4}. Dark regions mean low potential.}

 \end{figure}

In terms of the symplectic covering $\chi$ of Section \rref{Ssplcov}, the
dynamics can be described as follows. Given a Hamiltonian function
$H\in C^\infty(\pha)$, the lift $\chi^\ast H$ is a Hamiltonian
function on $\RR^4$. Let the curve $(x(t),p(t))$ be a solution of the
Hamiltonian equations associated with $\chi^\ast H$,
 \beq\label{GHaeqgeneral}
\dot p_j = - \,\frac{\partial(\chi^\ast H)}{\partial x^j}
 \,,~~~~~~
\dot x^j = \frac{\partial(\chi^\ast H)}{\partial p_j}
 \,,~~~~~~
j=1,2\,.
 \eeq
To be a solution is a local property. Since the map $\psi:\RR^4\to
T\times\mf t$ is a local symplectomorphism, then
$\psi\big((x(t),y(t)\big)$ is a solution of the Hamiltonian
equations of $\lambda^\ast H$ on $T\times\mf t$. According to
point 2 of Remark \rref{Rprojstrat}, this curve stays inside
$(T\times\mf t)_k$ for some $k=2,1,0$. Hence, $(x(t),p(t))$ stays
inside the corresponding $\RR^4_k$ and $\chi\big((x(t),p(t))\big)
= \chi_k\big((x(t),p(t))\big)$ is a curve in $\pha_k$. Since
$\chi_k$ is a local symplectomorphism by Theorem \rref{TphaR4},
then this curve is a solution of the Hamiltonian equations of the
Hamiltonian function $H|_{\pha_k}$ (restriction) on the stratum
$\pha_k$. This way, the Hamiltonian dynamics on $\pha$ w.r.t.\ $H$
is completely solved by the Hamiltonian dynamics w.r.t.\
$\chi^\ast H$ on $\RR^4$. Furthermore, the trajectories in $\cfg$
are given by $\pi\circ\chi\big((x(t),p(t))\big)$. Define
$\tilde\chi : \RR^2 \to \cfg$ to be the composition of the
covering $\vp:\RR^2\to T$, see \eqref{GcovT}, with the natural
projection $T\to\cfg$. Then $\pi\circ\chi\big((x(t),p(t))\big) =
\tilde\chi(x(t))$. Hence, for the discussion of the trajectories
in $\cfg$, it suffices to consider the trajectories $x(t)$ in
$\RR^2$. An explicit realization of the trajectories in $\cfg$ can
be obtained by passing to $\RR^2$ by means of the Hilbert map
$\rho_\cfg$, see \eqref{Gdefrhocfg}. One finds
$$
 \begin{array}{rcl}
\rho_\cfg\circ\tilde\chi(x(t)) & = &
 \left(
 \cos\big(
\frac{1}{\sqrt 6} x^1(t) + \frac{1}{\sqrt 2} x^2(t)
 \big)
+
 \cos\big(
\frac{1}{\sqrt 6} x^1(t) - \frac{1}{\sqrt 2} x^2(t)
 \big)
+
 \cos\big(
\sqrt{\frac{2}{3}} x^1(t)
 \big)
 \right.\,,
\\
 & &
 \phantom{\Big(}
 \left.
 \sin\big(
\frac{1}{\sqrt 6} x^1(t) + \frac{1}{\sqrt 2} x^2(t)
 \big)
+
 \sin\big(
\frac{1}{\sqrt 6} x^1(t) - \frac{1}{\sqrt 2} x^2(t)
 \big)
-
 \sin\big(
\sqrt{\frac{2}{3}} x^1(t)
 \big)
 \right)\,.
 \end{array}
$$
Now consider the Hamiltonian \eqref{GH}. One has
 \beq\label{GHR4}
 \begin{array}{c}
\chi^\ast H
 =
\frac{\delta^3}{2} (p_1^2 + p_2^2)
 +
\frac{1}{g^2\delta}
 \Big(3 -
 \cos\big(
\frac{x^1}{\sqrt 6} + \frac{x^2}{\sqrt 2}
 \big)
-
 \cos\big(
\frac{x^1}{\sqrt 6} - \frac{x^2}{\sqrt 2}
 \big)
-
 \cos\big(
\sqrt{\frac{2}{3}} x^1
 \big)
 \Big)
 \end{array}
 \eeq
The Hamiltonian has the standard structure, consisting of a
kinetic energy term and a potential term. The potential is
represented in Figure \rref{figpotential}. Its minimal value is $0$, it is taken
at the points
$$
 \begin{array}{c}
\big(\,3 l\,\sqrt{\frac{2}{3}}\,\pi\,,\,(3 l + 2 m)\,\sqrt 2\,\pi\,\big)\in\RR^2_0
 \,,~~~~
l,m\in\ZZ\,.
 \end{array}
$$
The maximal value is $\frac{1}{g^2\delta} \frac{9}{2}$, taken at
$$
 \begin{array}{c}
\big((3 l + 1)\,\frac{2}{\sqrt 3}\,\pi,(3 l + 2 m + 1)\,\sqrt 2\,\pi\big)
 \,,\,
\big((3 l + 2)\,\frac{2}{\sqrt 3}\,\pi,(3 l + 2 m + 2)\,\sqrt 2\,\pi\big)
 \in\RR^2_0
 \,,~~~~
l,m\in\ZZ\,.
 \end{array}
$$
In addition, the potential has saddle points at
$$
 \begin{array}{c}
\big(\,3 l\,\sqrt{\frac{2}{3}}\,\pi\,,\,(3 l + 2 m + 1)\,\sqrt
2\,\pi\,\big)\in\RR^2_1
 \,,~~~~~~
l,m\in\ZZ\,.
 \end{array}
$$
In the representation of $\RR^2$ in Figure \rref{figstrataR2}, the minima are
the points labelled by
$0$; they project to $\II\in\cfg_0$. The maxima are the points
labelled by $1$ and $2$, they project to the other two central elements
$\mr e^{i\frac{2}{3}\pi}\II$ and $\mr e^{i\frac{4}{3}\pi}\II\in\cfg_0$. The
saddle points are situated in the middle between points labelled by $1$ and $2$.

The Hamiltonian equations associated with $H$ are
 \beq\label{GeqsmoR4}
 \begin{array}[b]{rcl}
\dot p_1
 & = &
- \frac{1}{g^2\delta}\,\sqrt{\frac{2}{3}}\,
 \left(
\sin\big(\frac{1}{\sqrt 6} x^1\big)\,\cos\big(\frac{1}{\sqrt 2}\, x^2\big)
 +
\sin\big(\sqrt{\frac{2}{3}}\,x^1\big)
 \right)\,,
\\
\dot p_2
 & = &
- \frac{1}{g^2\delta}\,\sqrt 2\,
 \cos\big(\frac{1}{\sqrt 6} x^1\big)\,\sin\big(\frac{1}{\sqrt 2}\, x^2\big)\,,
\\
\dot x^j & = & \delta^3 p_j\,,~~~~~~j=1,2\,.
 \end{array}
 \eeq
Combining them, we obtain
 \beq\label{GeqsmoR4-2}
 \begin{array}[b]{rcl}
\ddot x^1
 +
\frac{\delta^2}{g^2}\,\sqrt{\frac{2}{3}}\,
 \left(
\sin\big(\frac{1}{\sqrt 6} x^1\big)\,\cos\big(\frac{1}{\sqrt 2}\, x^2\big)
 +
\sin\big(\sqrt{\frac{2}{3}}\,x^1\big)
 \right)
 & = & 0\,,
\\
\ddot x^2
 +
\frac{\delta^2}{g^2}\,\sqrt 2\,
 \cos\big(\frac{1}{\sqrt 6} x^1\big)\,\sin\big(\frac{1}{\sqrt 2}\, x^2\big)
 & = & 0\,.
 \end{array}
 \eeq
As mentioned above, this system of equations will be studied in
detail in a subsequent paper.
 \comment{
evtl. noch kurze Diskussion:

-- definiere $z$ und $w$

-- spezielle Loesungen $z=0$ und $w=0$, diese sind periodisch
(Pendel mit Frequenz ...)
 }%

Next, we comment on the discussion of the dynamics in terms of the invariants
of Section \rref{Sivr}. For a given Hamiltonian function $\tilde H\in
C^\infty(\RR^8)$, dynamics takes place on $\RR^8$ and is ruled by
the Poisson structure defined by the brackets of the coordinates
\eqref{GPoibraivr}. I.e., the equations of motion are given by
 \beq\label{Geqsmoivrgeneral}
\dot x_j = \{\tilde H,x_j\}
 \,,~~~~~~
(x_1,\dots,x_8) = (c_0,\dots,t_3)\,.
 \eeq
By construction of the Poisson structure, $\tpha$ is invariant
under the flow of $\tilde H$ for any $\tilde H\in
C^\infty(\RR^8)$. In terms of the invariants, the Hamiltonian \eqref{GH} reads
$$
 \begin{array}{c}
\tilde H = \frac{\delta^3}{2}\,t_2 +
\frac{1}{g^2\delta}\,(3-c_0)\,.
 \end{array}
$$
The second term corresponds to the potential term
in \eqref{GHR4}. Its level lines in $\tcfg$ are just straight
lines parallel to the $d_0$-axis, cf.\ Figure \rref{fighypocy}.
The minimum is at the corner $(c_0,d_0) = (3,0)$, the maxima are
at the corners $(c_0,d_0) = (-\frac 3 2,\pm\sqrt 3 \frac 3 2)$,
the saddle point is at the boundary point $(c_0,d_0) = (-1,0)$.

The corresponding equations of motion \eqref{Geqsmoivrgeneral} yield a highly
coupled system, which will not be reproduced here. At first sight it does not
seem to be easier to handle than the equations of motion in terms of the
symplectic covering \eqref{GeqsmoR4-2}. It will be a future task to study and
unravel this system.


\section*{Acknowledgements}


The authors would like to thank Sz.\ Charzynski, J.\ Huebschmann and I.P.\
Volobuev for helpful discussions on the invariants and the structure of the
reduced phase space.

M.\ S.\ acknowledges funding by the German Research Council (DFG) under project
nr.\ RU692/3.



\begin{thebibliography}{19}



\bibitem{AbraMar} Abraham, R.; Marsden, J.E.:
{\it Foundations of Mechanics.}
Addison-Wesley 1978

\bibitem{ArmsCushmanGotay}
Arms, J.M.; Cushman, R.H.; Gotay, M.J.:
A universal reduction procedure for Hamiltonian group actions.
In: {\it The geometry of Hamiltonian systems (Berkeley, CA, 1989)},
Math.~Sci.~Res.~Inst.~Publ.~22, Springer, New York, 1991, 33--51

\bibitem{configspace}
Charzy\'nski, S.; Kijowski, J.; Rudolph, G.; Schmidt, M.:
On the Stratified Classical Configuration Space of Lattice QCD.
J.\ Geom.\ Phys.\ {\bf 55} (2005) 137--178

\bibitem{CRS}
Charzy\'nski, S.; Rudolph, G.; Schmidt, M.:
On the Topology of the Reduced Classical Configuration Space of Lattice QCD.
hep-th/0512129

\bibitem{CuBa}
Cushman, R.H.; Bates, L.M.:
{\it Global Aspects of Classical Integrable Systems}.
Birkh\"auser-Verlag, Basel, 1997


\bibitem{Hue1}
Huebschmann, J.:
Kaehler spaces, nilpotent orbits, and singular
reduction. Mem.\ Amer.\ Math.\ Soc.\ {\bf 172} (2004) no.\ 814.
\smallskip

Huebschmann, J.:
Lie-Rinehart Algebras, Descent, and Quantization. In: {\it Galois theory,
Hopf algebras, and semiabelian categories}. Fields Inst.\ Commun., 43, Amer.\
Math.\ Soc., Providence, RI, 2004, pp. 295--316

\bibitem{Hue2}
Huebschmann, J.:
K\"ahler Quantization and Reduction, J.\ reine angew.\ Mathematik {\bf 591}
(2006) ({\tt math.sg/0207166})

\bibitem{Hue3}
Huebschmann, J.:
Classical phase space singularities and quantization.
4'th International Symposium ``Quantum Theory and Symmetries'', Varna,
2005. In: Quantum Theory and Symmetries IV. Ed. V.K.\ Dobrev, Heron
Press, Sofia 2006 (to appear).
\smallskip

Huebschmann, J.:
Singular Poisson-K\"ahler geometry of certain adjoint quotients.
In: Proceedings, The mathematical legacy of C. Ehresmann, Bedlewo, 2005,
Banach center publications (to appear)

\bibitem{Husemoeller}
Husemoller, D.:{\it Fibre Bundles.} McGraw-Hill 1966, Springer
1994

\bibitem{KiRu:JMP}
Kijowski, J.; Rudolph, G.:
Charge superselection sectors for qcd on the lattice.
J.\ Math.\ Phys.\ {\bf 46} (2005) 032303
\smallskip

Jarvis, P.D.; Kijowski, J.; Rudolph, G.: 
On the structure of the observable algebra of QCD on the lattice. 
J.\ Phys.\ A {\bf 38} (2005) 5359--5377

\bibitem{SGQ}
Landsman, N.P.; Pflaum, M.; Schlichenmaier, M.\ (eds.):
{\it Quantization of singular symplectic quotients.}
Progr.\ Math.\ {\bf 198}, Birkh\"auser 2001

\bibitem{LermanMontgomerySjamaar}
Lerman, E.; Montgomery, R.; Sjamaar, R.:
Examples of singular reduction.
In: {\it Symplectic geometry}, London Math. Soc. Lect. Note Ser. 192,
Cambridge Univ. Press, Cambridge 1993, pp. 127--155

\bibitem{OrtegaRatiu}
Ortega, J.P.; Ratiu, T.:
{\it Momentum maps and Hamiltonian reduction.}
Progress in Mathematics, 222, Birkh\"auser 2004

\bibitem{Perlmutter}
Perlmutter, M.; Rodriguez-Olmos, M.; Sousa-Dias, M.E.:
On the geometry of reduced cotangent bundles at zero momentum.
math.SG/0310437

\bibitem{Pflaum:LNM}
Pflaum, M.J.: Analytic and geometric study of stratified spaces.
Lect. Notes Math. 1768, Springer-Verlag, Berlin, 2001

\bibitem{Procesi}
Procesi, C.:
The invariant theory of $n\times n$ matrices.
Advances in Math. {\bf 19} (1976) 306--381

\bibitem{Schwarz:Smooth}
Schwarz, G.W.:
Smooth functions invariant under the action of a compact Lie group.
Topology {\bf 14} (1975) 63--68

\bibitem{SjamaarLerman}
Sjamaar, R.; Lerman, E.:
Stratified symplectic spaces and reduction.
Ann.\ of Math.\ (2) {\bf 134} (1991) 375--422

\bibitem{Weyl:Classical}
Weyl, H.: {\it The Classical Groups.} Princeton
University Press 1966, \S VIII.14


\end{thebibliography}
\end{document}